\documentclass[paper]{JFM-FLM_Au}

\makeatletter
\def\oddabsfooterflag{\hbox to \hsize{\hfil{\cppagefont\thepage}}}
\def\evenabsfooterflag{\hbox to \hsize{{\cppagefont\thepage}\hfil}}
\def\pagelimitfooter{\hbox to \textwidth{{\cppagefont\thepage}\hfil}}
\g@addto@macro\ps@headings{\def\@oddfoot{\hbox to \textwidth{\hfil{\cppagefont\thepage}}}}
\makeatother
\usepackage{amsmath,mathtools,bm}
\usepackage{xcolor}
\usepackage{graphicx}
\usepackage{enumitem}

\newtheorem{theorem}{Theorem}

\newcommand{\rb}{\bar\rho}\newcommand{\ub}{\bar u}\newcommand{\vb}{\bar v}\newcommand{\wb}{\bar w}\newcommand{\Tb}{\bar T}
\newcommand{\rh}{\hat\rho}\newcommand{\uh}{\hat u}\newcommand{\vh}{\hat v}\newcommand{\wh}{\hat w}\newcommand{\Th}{\hat T}
\definecolor{doilink}{rgb}{0.05,0.25,0.65}
\newcommand{\bibdoi}[1]{\href{https://doi.org/#1}{\textcolor{doilink}{doi:\nolinkurl{#1}}}}
\newcommand{\DX}{\mathcal D_X}
\newcommand{\D}{\mathcal D}
\newcommand{\kt}{k_t}
\newcommand{\kvec}{\mathbf{k}_t}
\newcommand{\vpar}{\hat v_\parallel}
\newcommand{\vperp}{\hat v_\perp}
\AtBeginDocument{\hypersetup{hidelinks}}

\newcommand{\Ky}{K_y}
\newcommand{\Kz}{K_z}
\newcommand{\Kt}{K_t}
\newcommand{\Om}{\Omega}

\lefttitle{V. Theofilis}
\righttitle{Journal of Fluid Mechanics}

\title{Rotational covariance, continuous spectrum and non-modal growth of three-dimensional planar weak viscous shock layers}

\author{Vassilis Theofilis}

\affiliation{The Stephen B. Klein Faculty of Aerospace Engineering, Technion, Israel Institute of
Technology, Haifa 3200003, Israel}

\corresau{Vassilis Theofilis, \email{vassilis@technion.ac.il}}

\begin{document}
\maketitle

\begin{abstract}
Three-dimensional linear stability of a weak shock layer is analyzed
with exact Reynolds-number-independent base-flow and stability equations, written in
shock viscous scales. For any base flow depending on the
shock-normal coordinate alone, with zero transverse velocity, the eigenvalue problem is proved
covariant under rotation of the transverse wavenumber vector. Its eigenvalues depend only on the
magnitude of this vector, and every eigenmode
decouples exactly into an in-plane acoustic--entropy branch, which carries all dilatation
and thermodynamic coupling, and an out-of-plane vortical branch, governed by a single
shear-diffusion equation. 
The continuous spectra are obtained using the 
Grosch--Salwen far-field companion matrices and,
independently, closed-form dispersion relations. The two agree
to machine precision.
The vortical continuous
spectrum retreats from the real axis exactly as the square of the transverse wavenumber, while
the acoustic--entropy continuous spectrum is strictly damped and, numerically, its least-damped point
does not increase with this wavenumber. Three-dimensional disturbances are therefore damped
at least as strongly as their two-dimensional counterparts with the same in-plane wavenumber,
as measured by the least-damped point of the spectrum, exactly for the vortical branch and numerically otherwise,
making the reduction the exact, Reynolds-free counterpart of Squire's transformation, in which
the wavenumber magnitude plays the role of the Reynolds number. 
Planar oblique shock stability results are obtained from normal-shock results by a Doppler shift.
Non-modal growth, measured in the Euclidean norm, is far stronger in the post-shock than in the pre-shock
gas.
Linear optimal disturbances of the shock-layer 
ripple the post-shock density field without displacing the shock front.
\end{abstract}


\section{Introduction}
\label{sec:intro}

The stability and response of a shock front treated as an infinitesimally thin
discontinuity has a long history, starting with \citet{Dyakov1954} and
\citet{Kontorovich1957}, who established the classical linear stability criterion for a
shock satisfying the Rankine--Hugoniot conditions and determined when such a discontinuity is
itself corrugationally unstable or spontaneously emits acoustic waves. \citet{Ribner1954} and
\citet{McKenzieWestphal1968} instead analyzed the transmission, reflection and
amplification of acoustic, entropy and vortical waves incident on an otherwise stable
discontinuity. \citet{DuckBalakumar1992} were the first to pose and solve the viscous linear stability
problem of the free shock layer, examining its own viscously diffused structure
\citep{GilbargPaolucci1953} rather than the corrugational stability of an idealized jump
or the scattering of external waves by one. Posing the problem in the shock's own
viscous scales, free of the Reynolds number $\Rey$ that enters explicitly in the
\textit{outer} compressible Navier--Stokes description, these authors found no unstable eigenmode within
the wavenumber and Mach number ranges examined. Rescaling the shock-normal coordinate, the
transverse wavenumbers and the eigenfrequency together by $\Rey$ absorbs it into these
same viscous scales and is shown herein to remove it identically, not merely
asymptotically, from the outer equations, recovering exactly an \textit{inner} form of the equations
which is valid at all Reynolds numbers. \citet{Humpherys2009} later showed, by asymptotic estimates and an extensive numerical Evans-function computation over a wide
range of specific-heat and viscosity-to-conductivity ratios, that the one-dimensional profile, without transverse variation,
is modally stable.

These analyses rest on the compressible Navier--Stokes--Fourier (NSF) equations applied to the
shock's own internal structure, a continuum description that is not universally valid but is
restricted to weak shocks at low Mach number. Numerical evidence supporting the NSF predicted
density and temperature shock profiles at low Mach numbers was provided in BGK-model kinetic solutions
\citep{LiepmannNarasimhaChahine1962} and direct simulation Monte Carlo (DSMC) solutions of
the underlying Boltzmann equation \citep{Bird1967}. Both classes of kinetic simulations reproduce the weak-shock 
layer NSF results, but systematically depart from them as the shock Mach number increases, a result 
confirmed experimentally by electron-beam density measurements \citep{Schmidt1969,Alsmeyer1976}.
Accordingly, the NSF description remains valid only while the
shock thickness stays large compared with the local mean free path, typically for
$M_1\lesssim2$ in a monatomic gas, although differences from kinetic shock structures grow with the
Mach number, first and most visibly on the upstream side of the layer \citep{LiepmannNarasimhaChahine1962}, and the shock is only about $3.4$ mean free paths
thick at $M_1=2$. The present analysis is therefore restricted to weak shock layers.

Two clarifications are in order in this context,
both bearing on the scope of the present work. 
First, the analysis of normal shocks discussed in what follows does not constitute a 
restriction, the reason being that for any steady, physically admissible shock layer the tangential
velocity components are constant across the entire layer \citep{Theofilis2026}, so an
arbitrary oblique shock reduces exactly to the normal-shock
problem treated below through a Galilean transformation to the frame moving with the tangential
velocity. This shifts the frequency of every disturbance by a constant real Doppler shift and leaves growth and damping rates unchanged. In other words, the normal-shock results that follow apply to
oblique planar shocks at the corresponding normal Mach number,
as discussed in detail in \S\ref{sec:theory-baseflow}. 
Second, the viscously diffuse shock layer is an exact, albeit numerically obtained, solution of the 
viscous laminar equations of motion in the continuum limit. Connections between the classical linear-interaction 
treatments of \citet{Ribner1954} and \citet{McKenzieWestphal1968} and the more recent inviscid analyses 
of \citet{DuckLasseigneHussaini}, for a shock attached to a wedge,
and of \citet{MilmanKarp2026}, for the normal shock ahead of a blunt body 
are worth exploring in future studies,
once the stability properties of a viscous shock layer have been established
in the realm of applicability of the continuum equations model. 
The viscously diffuse shock layer is the actual physical medium that an external disturbance must
traverse, and its receptivity and non-normal amplification properties, discussed in \S\ref{sec:pseudospectra},
are therefore of interest in this context.

In contrast with the compressible boundary layer, where a wall with the no-slip condition supports inviscid as well
as viscous instability mechanisms and a rich, multi-modal spectrum of unstable disturbances in
the compressible regime \citep{Mack1984}, the shock layer is unbounded on both sides, and its
purely diffusive steady laminar velocity and temperature profiles
connect two uniform states through the
Rankine--Hugoniot conditions. \citet{GroschSalwen1978} showed, for the Orr--Sommerfeld
equation on a semi-infinite domain, that whenever the coefficients of a stability operator tend
to constant values in the far field, the far-field behavior of every eigenmode is governed by
an algebraic companion-matrix system built from those limiting coefficients alone. This idea is exploited
in the present work, since it applies simultaneously to both sides of the layer, and the resulting companion
systems of \S\ref{sec:theory-gs} form the cornerstone of the present analysis.

At each Mach number and transverse wavenumber pair $(k_y,k_z)$ the companion systems map
the complex eigenfrequency plane into the far-field structure of the continuous spectrum of the shock
layer. For $k_t>0$, where the one-dimensional result of \citet{Humpherys2009} does not apply, discrete instability is excluded analytically in \S\ref{sec:theory-theorem-consequence} 
for the vortical branch and numerically for the acoustic--entropy branch, 
so the least-damped part of the spectrum is described by this far-field structure of the continuous
spectrum, a simplification to be discussed in what follows. 
Three-dimensional disturbances can be described by a combination of the transverse
wavenumbers, $k_t=(k_y^2+k_z^2)^{1/2}$, alongside a direction of propagation in the $(y,z)$ plane.
Two key questions are whether a rotational symmetry makes the problem depend on $k_t$
alone, rather than on $k_y$ and $k_z$ individually, and whether the acoustic, entropy and
vortical constituents of a disturbance can be addressed separately. Answers are provided by a
theorem\  whose rotational reduction is the exact,
Reynolds-free counterpart of the transformation of \citet{Squire1933} for the boundary layer,
stated and proven in \S\ref{sec:analytical}.
The present paper extends the recent Reynolds-dependent study of \citet{Theofilis2026} by recasting
the problem in the shock's own viscous scales, where the Reynolds number is eliminated exactly. Closed-form far-field
dispersion relations for both branches are presented, the continuous spectrum is characterized in mathematical
terms as \textit{essential}, and a
disturbance-energy bound is presented for the acoustic--entropy branch. Non-modal analyses of both the shock layer itself and its upstream cold
and downstream hot simplifications complete the analysis.

The paper is organized as follows. Section~\ref{sec:theory} derives the Reynolds-free base-flow and
stability equations from the Navier--Stokes--Fourier equations and the Gilbarg--Paolucci shock profile.
Section~\ref{sec:analytical} presents the main theoretical results in closed form, namely (i) the rotational covariance
and (ii) the decoupling theorem, (iii) an energy bound for the acoustic--entropy branch, (iv) the companion-matrix discriminant and the exact far-field dispersion relations and (v) the structure of the continuous spectrum.
Section~\ref{sec:results} verifies the formulation numerically and presents spectral maps, which are semi-analytic, since no discretization of the shock layer is involved.
Section~\ref{sec:pseudospectra} examines non-modal growth in the two far fields and its connection
to transient growth in the shock layer itself, through propagator and resolvent norms of the far-field generator
and Chebyshev collocation of the full operators. Appendices~\ref{app:evp-validation}--\ref{app:dispersion-check} verify the full-layer discretization, show that transient growth is bounded at the origin of the wavenumber plane and compare the exact dispersion relations with the discriminant maps. Conclusions are discussed in \S\ref{sec:conclusions}.

\section{Formulation}
\label{sec:theory}

\subsection{Inner vs. outer scales and the resulting basic flow}
\label{sec:theory-baseflow}

The compressible Navier--Stokes equations may be non-dimensionalized on the
upstream (cold-side) density, velocity, temperature and viscosity, and on an arbitrary upstream
\textit{outer} length scale $\delta_1^*$, resulting in the outer Reynolds
number $\Rey=\rho_1^*u_1^*\delta_1^*/\mu_1^*$. Alternatively, \citet{GilbargPaolucci1953} 
and \cite{DuckBalakumar1992} resolved the shock on its own natural, \textit{inner} viscous length scale
$\ell_{\mathrm{visc}}^*=\mu_1^*/(\rho_1^*u_1^*)=\delta_1^*/\Rey$.
The inner $(x,k_y,k_z,\omega)$ and outer $(X,\Ky,\Kz,\Om)$ 
streamwise coordinates, transverse wavenumbers and eigenvalues, as well as the spatial derivatives,
$\DX\equiv\frac{d}{dX}$ and $\D\equiv\frac{d}{dx}$ are related by the Reynolds number as follows
\begin{equation}
x=\Rey\,X,\qquad \DX=\Rey\,\D,\qquad
\Ky=\Rey\,k_y,\quad \Kz=\Rey\,k_z,\quad \Om=\Rey\,\omega.
\label{eq:rescaling}
\end{equation}
The full outer system, including the base-flow equations obtained at $O(1)$ and the linearized Navier-Stokes
equations in the form of the (modal) eigenvalue problem governing the small-amplitude perturbations 
have been presented and solved by \citet{Theofilis2026}. Continuity leads to
$\rb\ub=1$ in these units, while the bounded tangential components $\vb,\wb$ are shown
there to be constant for any physically admissible (bounded) steady layer. The general
oblique shock therefore reduces, without loss of generality, to a normal shock layer having
$\vb=\wb=0$, through a Galilean transformation to the frame moving with the tangential velocity $(\vb,\wb)$ of the oblique layer, under which $\omega=\omega'+k_y\vb+k_z\wb$, with $\omega'$ the frequency for the normal layer, that is a constant, real Doppler shift. Regarding the basic flow, elimination of density via
continuity leaves a coupled nonlinear ordinary differential equation system for the determination of $\ub,\Tb$, in which the Reynolds number 
appears only as the explicit coefficient of the viscous terms in these outer units.
Substituting \eqref{eq:rescaling} into the system of basic flow equations discussed by 
\citet{Theofilis2026} results in
\begin{subequations}
\label{eq:baseflow-inner}
\begin{align}
\left(\tfrac43\bar\mu+\bar\lambda\right)\D\ub
&= \ub+\frac{1}{\gamma M_1^2}\frac{\Tb}{\ub}-1-\frac{1}{\gamma M_1^2},\\[4pt]
\frac{\bar\mu}{\Pran\,(\gamma-1)M_1^2}\,\D\Tb
&= \frac{1}{\gamma(\gamma-1)M_1^2}\Tb-\frac12\ub^2+\left[1+\frac{1}{\gamma M_1^2}\right]\ub
-\frac{1}{(\gamma-1)M_1^2}-\frac12,
\end{align}
\end{subequations}
closed by the Rankine--Hugoniot conditions
\begin{subequations}
\label{eq:RH}
\begin{align}
u_1&=1,\qquad u_2=\frac{\gamma-1}{\gamma+1}+\frac{2}{(\gamma+1)M_1^2},\\
T_1&=1,\qquad T_2=\frac{\left(\gamma M_1^2-M_1^2+2\right)\left(1-\gamma+2\gamma M_1^2\right)}{(\gamma+1)^2M_1^2}.
\end{align}
\end{subequations}
System (\ref{eq:baseflow-inner}--\ref{eq:RH}),
from which shock layer profiles may be recovered at all
finite Reynolds numbers using the scaling (\ref{eq:rescaling}),
was put forward by \citet{GilbargPaolucci1953} 
and was also solved in the linear stability analysis of \citet{DuckBalakumar1992}.

The gas considered presently is monatomic and ideal with specific-heat ratio $\gamma=5/3$ and Prandtl number $\Pran=2/3$, and
$M_1$ is the upstream / cold-side Mach number. The dynamic viscosity follows the power law $\bar\mu(T)=T^{1/2}$ and the
bulk viscosity is $\bar\lambda=0$, as appropriate for a monatomic gas, so that the combinations
$\bar\lambda+\tfrac43\bar\mu$ and $\bar\lambda+\tfrac13\bar\mu$ of the momentum equations reduce to $\tfrac43\bar\mu$ and $\tfrac13\bar\mu$.

The inner scales can be used to recover the full shock layer physics. Dimensional
frequencies and transverse wavenumbers are related to their inner, non-dimensional
counterparts by the viscous length and time scales,
\begin{equation}
\omega^\ast=\frac{\omega}{t_{\rm visc}^\ast},\qquad
k_t^\ast=\frac{k_t}{\ell_{\rm visc}^\ast},\qquad\mathrm{where}\qquad
\ell_{\rm visc}^\ast=\frac{\mu_1^\ast}{\rho_1^\ast u_1^\ast},\qquad
t_{\rm visc}^\ast=\frac{\ell_{\rm visc}^\ast}{u_1^\ast},
\label{eq:dimensional-conversion}
\end{equation}
quantities fixed entirely by the upstream state. By contrast, in the outer formulation of \citet{Theofilis2026} 
the Reynolds number $\Rey=\rho_1^\ast u_1^\ast\delta_1^\ast/\mu_1^\ast$
depends on $\delta_1^\ast$, an outer length with no reference to the shock layer itself.
Two shock layers at the \emph{same} $M_1$ but with upstream densities in the ratio two, namely the same gas at two different ambient pressures
and fixed $T_1^\ast$, have viscous lengths in the ratio $\ell_{\rm visc}^{\ast(2)}=\tfrac12\ell_{\rm visc}^{\ast(1)}$,
since $\mu_1^\ast$ depends on temperature alone. In the present formulation, the \emph{same} inner wavenumber $k_t$ then corresponds to physical
transverse wavelengths $2\pi/k_t^\ast$ differing by a factor of two, a real, measurable difference that would be lost if the problem
were scaled on an arbitrarily chosen $\delta_1^\ast$. The spectrum, expressed in $(k_t,\omega)$, is therefore exactly invariant to that choice, while
\eqref{eq:dimensional-conversion} shows that, through the definitions of $\ell_{\rm visc}^\ast,t_{\rm visc}^\ast$, it responds to every actual change in the
upstream viscosity, density or velocity that set the shock thickness.

\subsection{Linear stability equations}
\label{sec:theory-lst}

When the equations of motion in outer variables $(X,\Ky,\Kz,\Om)$
are linearized about $\bar{\mathbf q}=(\rb,\ub,0,0,\Tb)^{\mathrm T}$ and the modal
ansatz $\widehat{\mathbf q}(X)\exp[\mathrm i(\Ky y+\Kz z-\Om t)]$ is introduced, the eigenvalue
problem (EVP) $\mathcal A\widehat{\mathbf q}=\mathrm i\Om\,\mathcal B\widehat{\mathbf q}$
results, as presented in eqs.~(3.4a)--(3.4e) of \citet{Theofilis2026}. Application of
the inner scaling \eqref{eq:rescaling} to those equations converts them into the
Reynolds-free linearized stability equations, written below in the inner variables $(x,k_y,k_z,\omega)$.

\vspace{6pt}

\begingroup
\footnotesize
\setlength{\jot}{0pt}
\setlength{\abovedisplayskip}{1pt plus 1pt minus 1pt}
\setlength{\belowdisplayskip}{1pt plus 1pt minus 1pt}
\begin{subequations}
\label{eq:lnse}

\noindent\textbf{Continuity:}\vspace{-2pt}
\begin{equation}
(\ub\D+\D\ub+\mathrm ik_y\vb+\mathrm ik_z\wb)\,\rh+(\rb\D+\D\rb)\,\uh
+\mathrm ik_y\rb\,\vh+\mathrm ik_z\rb\,\wh=\mathrm i\omega\rh.
\label{eq:lnse-cont}
\end{equation}

\vspace{6pt}

\noindent\textbf{$x$-momentum:}\vspace{-2pt}
\begin{align}
&\left(\rh\,\ub\D\ub+\frac{1}{\gamma M_1^2}\D(\rh\Tb)\right)\notag\\
&\quad+\left[\rb(\ub\D+\mathrm ik_y\vb+\mathrm ik_z\wb+\D\ub)
-\left((\bar\lambda+\tfrac43\bar\mu)\D^2+\Big(\tfrac{d\bar\lambda}{dT}\D\Tb+\tfrac43\tfrac{d\bar\mu}{dT}\D\Tb\Big)\D-\bar\mu(k_y^2+k_z^2)\right)\right]\uh\notag\\
&\quad-\mathrm ik_y\left((\bar\lambda+\tfrac13\bar\mu)\D-\tfrac23\tfrac{d\bar\mu}{dT}\D\Tb+\tfrac{d\bar\lambda}{dT}\D\Tb\right)\vh
-\mathrm ik_z\left((\bar\lambda+\tfrac13\bar\mu)\D-\tfrac23\tfrac{d\bar\mu}{dT}\D\Tb+\tfrac{d\bar\lambda}{dT}\D\Tb\right)\wh\notag\\
&\quad+\Bigg[\frac{1}{\gamma M_1^2}\D\rb-\mathrm ik_y\frac{d\bar\mu}{dT}\D\vb-\mathrm ik_z\frac{d\bar\mu}{dT}\D\wb
-\left(\frac{d^2\bar\lambda}{dT^2}+\frac43\frac{d^2\bar\mu}{dT^2}\right)(\D\Tb)(\D\ub)\notag\\
&\hspace{0.4em}-\left(\frac{d\bar\lambda}{dT}+\frac43\frac{d\bar\mu}{dT}\right)\D^2\ub\Bigg]\Th
+\left[\frac{\rb}{\gamma M_1^2}-\left(\frac{d\bar\lambda}{dT}+\frac43\frac{d\bar\mu}{dT}\right)\D\ub\right]\Th'
=\mathrm i\omega\rb\uh.
\label{eq:lnse-xmom}
\end{align}

\vspace{6pt}

\noindent\textbf{$y$-momentum:}\vspace{-2pt}
\begin{align}
&\left(\rh\,\ub\D\vb+\frac{\mathrm ik_y\Tb}{\gamma M_1^2}\rh\right)
+\left[\rb\D\vb-\mathrm ik_y\left((\bar\lambda+\tfrac13\bar\mu)\D+\frac{d\bar\mu}{dT}\D\Tb\right)\right]\uh\notag\\
&\quad+\left[\rb(\ub\D+\mathrm ik_y\vb+\mathrm ik_z\wb)
-\left(\bar\mu\D^2+\frac{d\bar\mu}{dT}\D\Tb\cdot\D-\bar\mu(k_y^2+k_z^2)-\bar\lambda k_y^2-\tfrac13\bar\mu k_y^2\right)\right]\vh
+\Big[k_yk_z(\bar\lambda+\tfrac13\bar\mu)\Big]\wh\notag\\
&\quad+\left[\frac{\mathrm ik_y\rb}{\gamma M_1^2}-\mathrm ik_y\frac{d\bar\lambda}{dT}\D\ub
+\tfrac23\mathrm ik_y\frac{d\bar\mu}{dT}\D\ub-\frac{d^2\bar\mu}{dT^2}(\D\Tb)(\D\vb)-\frac{d\bar\mu}{dT}\D^2\vb\right]\Th
-\frac{d\bar\mu}{dT}\D\vb\,\Th'
=\mathrm i\omega\rb\vh.
\label{eq:lnse-ymom}
\end{align}

\vspace{6pt}

\noindent\textbf{$z$-momentum:}\vspace{-2pt}
\begin{align}
&\left(\rh\,\ub\D\wb+\frac{\mathrm ik_z\Tb}{\gamma M_1^2}\rh\right)
+\left[\rb\D\wb-\mathrm ik_z\left((\bar\lambda+\tfrac13\bar\mu)\D+\frac{d\bar\mu}{dT}\D\Tb\right)\right]\uh
+\Big[k_yk_z(\bar\lambda+\tfrac13\bar\mu)\Big]\vh\notag\\
&\quad+\left[\rb(\ub\D+\mathrm ik_y\vb+\mathrm ik_z\wb)
-\left(\bar\mu\D^2+\frac{d\bar\mu}{dT}\D\Tb\cdot\D-\bar\mu(k_y^2+k_z^2)-\bar\lambda k_z^2-\tfrac13\bar\mu k_z^2\right)\right]\wh\notag\\
&\quad+\left[\frac{\mathrm ik_z\rb}{\gamma M_1^2}-\mathrm ik_z\frac{d\bar\lambda}{dT}\D\ub
+\tfrac23\mathrm ik_z\frac{d\bar\mu}{dT}\D\ub-\frac{d^2\bar\mu}{dT^2}(\D\Tb)(\D\wb)-\frac{d\bar\mu}{dT}\D^2\wb\right]\Th
-\frac{d\bar\mu}{dT}\D\wb\,\Th'
=\mathrm i\omega\rb\wh.
\label{eq:lnse-zmom}
\end{align}

\vspace{6pt}

\noindent\textbf{Energy:}\vspace{-2pt}
\begin{align}
&\left(\ub\D\Tb+(\gamma-1)\Tb\D\ub\right)\rh
+\Big[\rb\D\Tb+(\gamma-1)\rb\Tb\D\notag\\
&\hspace{0.4em}
-\gamma(\gamma-1)M_1^2\big((2\bar\lambda+\tfrac83\bar\mu)(\D\ub)\D+2\bar\mu(\D\vb)\mathrm ik_y+2\bar\mu(\D\wb)\mathrm ik_z\big)\Big]\uh\notag\\
&\quad+\left[\mathrm ik_y(\gamma-1)\rb\Tb-2\mathrm ik_y\gamma(\gamma-1)M_1^2\bar\lambda(\D\ub)
+\tfrac43\mathrm ik_y\gamma(\gamma-1)M_1^2\bar\mu(\D\ub)-2\gamma(\gamma-1)M_1^2\bar\mu(\D\vb)\D\right]\vh\notag\\
&\quad+\left[\mathrm ik_z(\gamma-1)\rb\Tb-2\mathrm ik_z\gamma(\gamma-1)M_1^2\bar\lambda(\D\ub)
+\tfrac43\mathrm ik_z\gamma(\gamma-1)M_1^2\bar\mu(\D\ub)-2\gamma(\gamma-1)M_1^2\bar\mu(\D\wb)\D\right]\wh\notag\\
&\quad+\Bigg[\rb(\ub\D+\mathrm ik_y\vb+\mathrm ik_z\wb)+(\gamma-1)\rb\D\ub\notag\\
&\hspace{0.4em}-\frac{\gamma}{\Pran}\Big(\bar\mu\D^2+2\frac{d\bar\mu}{dT}\D\Tb\cdot\D-\bar\mu(k_y^2+k_z^2)\Big)
-\frac{\gamma}{\Pran}\left(\frac{d^2\bar\mu}{dT^2}(\D\Tb)^2+\frac{d\bar\mu}{dT}\D^2\Tb\right)\notag\\
&\hspace{0.4em}-\gamma(\gamma-1)M_1^2\Big[\big(\tfrac43\tfrac{d\bar\mu}{dT}+\tfrac{d\bar\lambda}{dT}\big)(\D\ub)^2
+\tfrac{d\bar\mu}{dT}(\D\vb)^2+\tfrac{d\bar\mu}{dT}(\D\wb)^2\Big]\Bigg]\Th\notag\\
&= \mathrm i\omega\rb\Th.
\label{eq:lnse-energy}
\end{align}
\end{subequations}
\endgroup

 In \eqref{eq:lnse} a bar denotes a base-flow quantity, i.e. a steady function of $x$ alone, and a hat the corresponding perturbation amplitude function. Equations are written in a form collecting the coefficients of the unknown perturbation vector,
$\hat{\mathbf q}=(\rh,\uh,\vh,\wh,\Th)^\mathrm{T}$, and can be written in compact form as the generalized eigenvalue problem
\begin{equation}
\label{eqn:LST_EVP}
\mathbf L\hat{\mathbf q}=\mathrm i\omega\mathbf B\hat{\mathbf q},
\end{equation}
with $\mathbf B=\mathrm{diag}(1,\rb,\rb,\rb,\rb)$. A prime applied to a hatted quantity, as in $\Th'$, denotes $\D\Th$ promoted to an independent unknown of the first-order companion system of \S\ref{sec:theory-gs} so that, e.g., $\Th$ and $\Th'$ may be weighted by different coefficients within the same row, \eqref{eq:lnse-xmom}.

Equation \eqref{eqn:LST_EVP} describes modal stability, in which the Fourier ansatz $\exp(-\mathrm i\omega t)$ has been imposed on the time dependence. If no such assumption is made and the temporal derivative is retained, the linearized equations \eqref{eq:lnse} become the initial-value problem
\begin{equation}
\label{eqn:LST_IVP}
\frac{\partial\hat{\mathbf q}}{\partial t}=\mathbf A\hat{\mathbf q},\qquad \mathbf A=-\mathbf B^{-1}\mathbf L,
\end{equation}
whose eigenvalues $-\mathrm i\omega$ recover the modal problem \eqref{eqn:LST_EVP}, while its solution $\hat{\mathbf q}(t)=\exp(\mathbf A t)\,\hat{\mathbf q}(0)$ also contains the non-modal, transient growth of \S\ref{sec:pseudospectra} that eigenvalues alone cannot describe.

\subsection{Grosch--Salwen far-field companion systems}
\label{sec:theory-gs}

The Reynolds-free system \eqref{eq:lnse}
has constant coefficients in the far field, $x\to\pm\infty$, where the base-flow variables $\bar\rho,\bar u,\bar T$
assume uniform values and every $x$-derivative vanishes. The resulting system can be solved by the method
discussed in the classical work of \citet{GroschSalwen1978} for the Orr--Sommerfeld equation. In the far-field limits of the
present configuration \eqref{eq:lnse} collapses into two independent companion-matrix systems, 
a $7\times7$ in-plane acoustic--entropy system for
$\hat{\mathbf q}=(\rh,\uh,\D\uh,\vh,\D\vh,\Th,\D\Th)^\mathrm{T}$ and a $2\times2$ out-of-plane vortical system for
$\hat{\mathbf q}=(\wh,\D\wh)^\mathrm{T}$. The vorticity system follows from \eqref{eq:lnse-zmom} at
$k_z=0$, $\bar v\equiv\bar w\equiv0$,
\begin{equation}
\rho_\infty u_\infty\,\D\wh-\mu_\infty\,\D^2\wh+\mu_\infty k_y^2\,\wh
-\mathrm i\omega\rho_\infty\,\wh=0.
\label{eq:gs-vorticity}
\end{equation}
The acoustic--entropy system reduces to the far-field continuity, streamwise-momentum,
transverse-momentum and energy equations
\begin{subequations}
\label{eq:gs-at}
\begin{align}
u_\infty\D\rh+\rho_\infty\D\uh+\mathrm ik_y\rho_\infty\vh&=\mathrm i\omega\rh,\\[4pt]
\frac{T_\infty}{\gamma M_1^2}\D\rh+\rho_\infty u_\infty\D\uh
-\left(\lambda_\infty+\tfrac43\mu_\infty\right)\D^2\uh+\mu_\infty k_y^2\uh\notag\\
\qquad-\mathrm ik_y\left(\lambda_\infty+\tfrac13\mu_\infty\right)\D\vh
+\frac{\rho_\infty}{\gamma M_1^2}\D\Th&=\mathrm i\omega\rho_\infty\uh,\\[4pt]
\mathrm ik_y\frac{T_\infty}{\gamma M_1^2}\rh
-\mathrm ik_y\left(\lambda_\infty+\tfrac13\mu_\infty\right)\D\uh
+\rho_\infty u_\infty\D\vh-\mu_\infty\D^2\vh\notag\\
\qquad+\left(\lambda_\infty+\tfrac43\mu_\infty\right)k_y^2\vh
+\mathrm ik_y\frac{\rho_\infty}{\gamma M_1^2}\Th&=\mathrm i\omega\rho_\infty\vh,\\[4pt]
(\gamma-1)\rho_\infty T_\infty\D\uh+\mathrm ik_y(\gamma-1)\rho_\infty T_\infty\vh
+\rho_\infty u_\infty\D\Th\notag\\
\qquad-\frac{\gamma\mu_\infty}{\Pran}\D^2\Th
+\frac{\gamma\mu_\infty}{\Pran}k_y^2\Th&=\mathrm i\omega\rho_\infty\Th.
\end{align}
\end{subequations}
Systems \eqref{eq:gs-vorticity} and \eqref{eq:gs-at} are the Reynolds-free counterparts of
the curves computed at fixed Reynolds number values by \citet{Theofilis2026}. Here, for each $k_y$, a single solution per far-field
state at fixed $M_1,\gamma,\Pran$ determines the branch structure at all Reynolds numbers.

\section{Analytical results}
\label{sec:analytical}

This section presents results obtained in closed form from analysis of
the systems \eqref{eq:lnse} and (\ref{eq:gs-vorticity}--\ref{eq:gs-at}), namely
the rotational covariance and
decoupling theorem for the acoustic--entropy and vortical branches in \S\ref{sec:theory-theorem}, 
a disturbance-energy bound for the acoustic--entropy branch in \S\ref{sec:theory-energy-ae}, 
the companion-matrix discriminant in \S\ref{sec:theory-companion-discriminant}, 
the equivalent formulation as exact far-field dispersion relations of the
two branches in \S\ref{sec:analytical-dispersion} and the characterization of the
spectrum as essential spectrum in \S\ref{sec:theory-essential-spectrum}.

\subsection{Rotational covariance and exact decoupling of the continuous spectrum}
\label{sec:theory-theorem}

\begin{theorem}[SO(2) covariance of the Reynolds-free transverse EVP\footnote{An exact restatement, in the Reynolds-free inner variables $(x,k_y,k_z,\omega)$, of Theorem~1 of \citet{Theofilis2026}.}]
\label{thm:main}
If the Reynolds-free laminar base state of a shock layer has the form $(\rb(x),\ub(x),0,0,\Tb(x))^\mathrm{T}$,
the eigenvalue $\omega$ of the corresponding three-dimensional linear stability equations
depends on $(k_y,k_z)$ only through $\kt$, and every three-dimensional eigenmode is
exactly one of two kinds: either an in-plane \textbf{acoustic--entropy (AE)} eigenmode at
$k_z=0,\,k_y=\kt$, with $\vpar\equiv\vh$, having an eigenfunction
$(\rh,\uh,\vpar,\Th)$ in which the streamwise velocity $\uh$ is fully coupled and
carries dilatation with $\vpar$; or an out-of-plane \textbf{vortical (V)} eigenmode at
$k_z=0,\,k_y=\kt$, with $\vperp\equiv\wh$ having an eigenfunction $\vperp$ alone, governed
by eq.~\eqref{eq:vorticity-Xspace}.
\end{theorem}

The covariance theorem is proved in two steps:
\S\ref{sec:theory-theorem-branches} identifies the two branches and their governing
equations, and \S\ref{sec:theory-theorem-separability} establishes the exact
decoupling and rotational covariance that complete the proof.

\subsubsection{The acoustic--entropy and vortical branches}
\label{sec:theory-theorem-branches}

Small-amplitude disturbances of a compressible fluid otherwise at rest in a spatially
uniform medium have been classified by \citet{Kovasznay1953}
into three independent constituents: an acoustic mode carrying the entire pressure and
dilatation, a passively advected entropy (temperature) mode, and a solenoidal vortical
mode. It is interesting to examine whether the continuous spectrum of the shock layer
admits an analogous separation.

Since the shock layer is unbounded and \eqref{eq:lnse} reduces to the far-field companion
systems of \S\ref{sec:theory-gs} as $x\to\pm\infty$, at each real $(k_y,k_z)$ it supports a
continuum of neutral or damped eigenmodes, not a discrete set of isolated poles.
Furthermore, \citet{Humpherys2009} showed, by asymptotic estimates and extensive numerical Evans-function computations, that no
discrete unstable eigenvalue exists\  for the one-dimensional ($\kt=0$) profile.
The total transverse wavenumber and the rotated transverse velocity
components are introduced as
\begin{equation}
\kt \equiv \sqrt{k_y^2+k_z^2}, \qquad
\vpar \equiv \frac{k_y\vh+k_z\wh}{\kt}, \qquad
\vperp \equiv \frac{-k_z\vh+k_y\wh}{\kt},
\label{eq:rotation-inner}
\end{equation}
defining an orthogonal rotation of the perturbation vector $(\vh,\wh)$ for
$(k_y,k_z)\neq(0,0).$\footnote{identical in form to its outer counterpart, since
$\Ky/\Kt=k_y/\kt$ is scale-invariant under \eqref{eq:rescaling} ($\Kt\equiv\Rey\,\kt$
cancels the same $\Rey$ as $\Ky,\Kz$).}

In this section it is shown that, for a base flow having the form $\bar{\mathbf{q}}=(\rb(x),\ub(x),0,0,\Tb(x))^\mathrm{T}$,
the continuous spectrum decouples exactly into two families, and the eigenvalue $\omega$ depends on the
wavenumber pair $(k_y,k_z)$ only through their magnitude $\kt$.

The first family is the in-plane compressible \textit{acoustic--entropy} (AE) branch at $k_z=0,\,k_y=\kt$, which carries
$\uh$ together with the in-plane transverse velocity $\vpar\equiv\vh$ and has the eigenfunction $(\rh,\uh,\vpar,\Th)^{\mathrm T}$.
The second family is the out-of-plane, purely solenoidal \textit{vortical} (V) branch, at the same $k_z=0,\,k_y=\kt$, with
$\vperp\equiv\wh$ alone. It is governed by the $k_z=0,\bar w\equiv0$
reduction of eq.~\eqref{eq:lnse-zmom}, a universal, $\Rey$-free equation for
$\wh(x)$ at fixed $M_1,\gamma,\Pran,k_y$,
\begin{equation}
\rb\ub\,\D\wh-\frac{d\bar\mu}{dT}\D\Tb\,\D\wh-\bar\mu\,\D^2\wh
+\bar\mu k_y^2\,\wh-\mathrm i\omega\rb\,\wh=0,
\label{eq:vorticity-Xspace}
\end{equation}
in which $k_y^2$, hence $\kt^2$ at $k_y=\kt$, enters nowhere else. Together with the
$\kt$-only dependence of $\omega$ established next, this is the exact decoupling asserted
by the covariance theorem above.

It is noted that $\Th'\equiv\D\Th$ appears explicitly in the AE family as an unknown alongside $\Th$,
since the momentum equation \eqref{eq:lnse-xmom} multiplies $\Th$ and $\D\Th$
by distinct coefficients through the temperature-dependent transport properties and the equation of state.
The block therefore solves for both simultaneously, tied only by the definitional relation $\D\Th=\Th'$.
$\Th$ itself remains fully determined, no differently than if $\Th'$ had been eliminated in favor of
$\D\Th$ throughout.

\subsubsection{Spectrum separability}
\label{sec:theory-theorem-separability}

Spectrum separability follows because $\bar v=\bar w=0$ removes every base-flow Doppler
and cross-viscosity term proportional to the base flow transverse velocity components from the linearized equations,
splitting what remains into two independent blocks: a \textit{scalar acoustic--entropy block} built from the continuity, $x$-momentum and energy equations,
\eqref{eq:lnse-cont}, \eqref{eq:lnse-xmom} and \eqref{eq:lnse-energy}, respectively, and
a \textit{vector transverse-momentum block} composed of the two transverse-momentum equations,
\eqref{eq:lnse-ymom} and \eqref{eq:lnse-zmom}.
The \emph{scalar} and \emph{vector} labels describe behavior under the rotation
\eqref{eq:rotation-inner} itself: $\uh,\vh,\wh$ are not treated as components of
the velocity vector but grouped by how each transforms under this rotation of the
transverse $(y,z)$ plane about the $x$-axis, which leaves the $x$-direction fixed, so
$\uh$, lying along that axis, is invariant, while $\vh,\wh$, perpendicular to it,
transform together.

Inspection of the scalar block equations shows that
$\vh$ and $\wh$ appear only through the combination $\mathrm
ik_y\vh+\mathrm ik_z\wh$, with every other coefficient depending on $(k_y,k_z)$ only
through $\kt^2\equiv k_y^2+k_z^2$, and the same combination is the only way in which $\vh$ and $\wh$ enter the energy equation.
The vector block follows the same $\kt^2$ dependence, but its two rows also carry
an off-diagonal term $k_yk_z(\bar\lambda+\tfrac13\bar\mu)$ which provides the only coupling between $\vh$ and $\wh$ 
whenever $k_z\neq0$. The rotation \eqref{eq:rotation-inner} removes this coupling identically for every
wavenumber vector by aligning one transverse axis with $\kvec$ and the other
perpendicular to it. Along the aligned component $\vpar$, the viscous diffusion carries
the longitudinal combination $\bar\lambda+\tfrac43\bar\mu$; along the perpendicular component
$\vperp$, it carries only shear viscosity $\bar\mu$:
\begin{subequations}
\label{eq:diag-inner}
\begin{equation}
-\Big(\bar\mu\D^2+\frac{d\bar\mu}{dT}\D\Tb\D\Big)+(\bar\lambda+\tfrac43\bar\mu)\kt^2\ \ \text{acting on }\vpar,
\label{eq:diag-inner-par}
\end{equation}
\begin{equation}
-\Big(\bar\mu\D^2+\frac{d\bar\mu}{dT}\D\Tb\D\Big)+\bar\mu\kt^2\ \ \text{acting on }\vperp,
\label{eq:diag-inner-perp}
\end{equation}
\end{subequations}
now fully decoupled from one another. The same rotation collapses the scalar block's
coupling to $k_y\vh+k_z\wh=\kt\,\vpar$ identically, with no remaining $\vperp$-dependence.

Assembled in the rotated variables, the scalar block together with the $\vpar$ component
reproduces exactly the original acoustic--entropy system at
$k_z=0,\,k_y=\kt$; the now-unforced $\vperp$ component reproduces exactly the vortical
equation, \eqref{eq:lnse-zmom} at the same point, eq.~\eqref{eq:vorticity-Xspace}. Because this rotation
is invertible for every nonzero wavenumber vector, it matches every oblique
three-dimensional eigenmode at $(k_y,k_z)$ to exactly one pair of eigenmodes of these two
decoupled problems at $\kt$, and conversely, which is exactly the covariance and
two-branch decoupling asserted by the covariance theorem. \hfill\fullsquare

In the rotated frame, the two branches are exactly the far-field companion systems
\eqref{eq:gs-vorticity} and \eqref{eq:gs-at} of \S\ref{sec:theory-gs} in the uniform limit
$x\to\pm\infty$,\footnote{The same exact decoupling and covariance hold, unchanged in form,
for the Reynolds-dependent outer system of \citet{Theofilis2026}. The rescaling
\eqref{eq:rescaling} leaves the transverse wavenumber ratio $\Ky/\Kt=k_y/\kt$ invariant, so
the rotation \eqref{eq:rotation-inner} carries over directly to the outer variables
$(\Ky,\Kz)$.} distinguished by dilatation. The in-plane
branch $(\rh,\uh,\vpar,\Th)$ is \emph{compressible}, since the original dilatation
contribution $\mathrm ik_y\vh+\mathrm ik_z\wh$ collapses to $\mathrm i\kt\,\vpar$. Together
with the equation of state coupling $\rh,\Th$ to the pressure perturbation, this block
supports a genuine compressible, pressure-bearing disturbance, hence the qualification of the disturbance as \emph{acoustic}.
On the other hand, the base-flow gradients ($\D\rb,\D\ub,\D\Tb,d\bar\mu/dT$) couple density,
velocity and temperature, such that the scalar block cannot be split further into a
separate pressure-carrying acoustic pair and a pressure-free entropy/temperature mode
as in the classical \citet{Kovasznay1953} decomposition (valid only in a spatially uniform
medium). Consequently, this branch of the continuous spectrum is termed \emph{acoustic--entropy}.
The out-of-plane eigenvalue-spectrum branch is defined by $\vperp$ alone. By construction
it carries no dilatation, does not appear in continuity or the equation of state, and
cannot generate density, pressure or temperature fluctuation of its own. All that remains
governing this perturbation is the scalar shear-diffusion equation
\eqref{eq:vorticity-Xspace}: decoupled from every thermodynamic variable and carrying zero
dilatation, $\vperp$ is precisely a \emph{vortical} (vorticity) mode.

\subsubsection{Consequence for shock-layer stability}
\label{sec:theory-theorem-consequence}

The covariance theorem has an important consequence for shock-layer
stability. The diffusive terms in eq.~\eqref{eq:diag-inner} grow explicitly with $\kt^2$,
with coefficient $(\bar\lambda+\tfrac43\bar\mu)$ along $\vpar$ and $\bar\mu$ along $\vperp$, while every
other term is unchanged. Since $\bar\lambda+\tfrac43\bar\mu,\bar\mu>0$ and $\kt^2=k_y^2+k_z^2$ grows
monotonically with $|k_z|$ at fixed $k_y$, this suggests that increasing $\kt$, i.e.\ switching on
$k_z\neq0$, increases the damping. The eigenfunctions also change with $\kt$, however, so the expectation needs proof,
which can be made exact for the purely scalar vortical branch, as follows.
 
Writing eq.~\eqref{eq:vorticity-Xspace} as ${\cal L}\vperp=\mathrm i\omega\rb\vperp$ with
\begin{equation}
{\cal L}\equiv\rb\ub\D-\Big(\bar\mu\D^2+\frac{d\bar\mu}{dT}\D\Tb\D\Big)+\bar\mu\kt^2
=\rb\ub\D-\D(\bar\mu\D\,\cdot\,)+\bar\mu\kt^2,
\label{eq:operator-def}
\end{equation}
the diffusive part is recognized as the standard divergence-form (Sturm--Liouville)
operator $-\D(\bar\mu\D\,\cdot\,)$. The inner product over the shock-normal
coordinate is defined by
\begin{equation}
\langle f,g\rangle\equiv\int_{-\infty}^{\infty}f^\ast g\,\mathrm dx.
\label{eq:inner-product-def}
\end{equation}
Forming the inner product of $\vperp$ with both sides of eigenvalue equation
\eqref{eq:vorticity-Xspace} gives the Rayleigh quotient
\begin{equation}
\mathrm i\omega=\frac{\langle\vperp,{\cal L}\vperp\rangle}{\langle\vperp,\rb\vperp\rangle},
\label{eq:rayleigh-quotient-compact}
\end{equation}
in which the denominator, $\langle\vperp,\rb\vperp\rangle=\int_{-\infty}^{\infty}\rb|\vperp|^2\,\mathrm dx$,
is real and strictly positive since $\bar\rho>0$ is a physical density. Expanding the
numerator and integrating the diffusive part once by parts, its boundary contribution
vanishing for an eigenmode that decays as $|x|\rightarrow\infty$, results in
\begin{equation}
\mathrm i\omega\!\int_{-\infty}^{\infty}\!\!\rb|\vperp|^2\,\mathrm dx
=\int_{-\infty}^{\infty}\!\!\vperp^\ast\rb\ub\,\D\vperp\,\mathrm dx
+\int_{-\infty}^{\infty}\!\!\bar\mu|\D\vperp|^2\,\mathrm dx
+\kt^2\!\int_{-\infty}^{\infty}\!\!\bar\mu|\vperp|^2\,\mathrm dx,
\label{eq:rayleigh-quotient}
\end{equation}
with $\bar\mu(x)$ kept inside both diffusive integrals throughout, since it
varies across the shock. This contrasts with the analogous
acoustic--entropy identity of \S\ref{sec:theory-energy-ae}, which is built on the
far-field system \eqref{eq:gs-at} alone. The advection integral on the right hand side is complex in general, while
the diffusion and $\kt^2$ integrals are real, since $\bar\mu(x)>0$ throughout.
Equating real and imaginary parts of \eqref{eq:rayleigh-quotient} separates
$\omega=\omega_r+\mathrm i\omega_i$ exactly, with $\int\rb|\vperp|^2\,\mathrm dx>0$
throughout:
\begin{subequations}
\label{eq:rayleigh-quotient-split}
\begin{align}
\omega_r\int\rb|\vperp|^2\,\mathrm dx&=\mathrm{Im}\Big(\int\vperp^\ast\rb\ub\,\D\vperp\,\mathrm dx\Big),
\label{eq:rayleigh-quotient-split-r}\\
-\,\omega_i\int\rb|\vperp|^2\,\mathrm dx&=\mathrm{Re}\Big(\int\vperp^\ast\rb\ub\,\D\vperp\,\mathrm dx\Big)
+\int\bar\mu|\D\vperp|^2\,\mathrm dx+\kt^2\int\bar\mu|\vperp|^2\,\mathrm dx.
\label{eq:rayleigh-quotient-split-i}
\end{align}
\end{subequations}
In the above system the wavenumber $\kt^2$ appears explicitly only in
\eqref{eq:rayleigh-quotient-split-i}, as a real, non-negative term growing with $\kt$.
This does not by itself make $\omega_r$ independent of $\kt$, nor $\omega_i$ monotone in
$\kt$: the eigenfunction $\vperp$ itself depends on $\kt$, and since the kinematic viscosity
$\bar\mu/\rb$ varies across the layer, the $\kt^2$ term is not a multiple of the mass term
$\rb|\vperp|^2$ and does not simply shift the spectrum. What can be established for discrete eigenvalues of the full operator is a
bound, derived next.

A discrete eigenvalue of this branch is, by definition, carried by a genuinely decaying
eigenfunction, $\vperp,\D\vperp\to0$ as $x\to\pm\infty$, and for such a mode the advection
integral in \eqref{eq:rayleigh-quotient-split-i} is in fact \emph{purely imaginary}, so its
real part vanishes identically rather than merely being left unevaluated. Integrating by
parts once more and using the exact base-flow mass-flux conservation $\D(\rb\ub)=0$
(\S\ref{sec:theory-baseflow}) gives
\begin{equation}
\int_{-\infty}^{\infty}\vperp^\ast\rb\ub\,\D\vperp\,\mathrm dx
=-\int_{-\infty}^{\infty}(\D\vperp)^\ast\rb\ub\,\vperp\,\mathrm dx
=-\left(\int_{-\infty}^{\infty}\vperp^\ast\rb\ub\,\D\vperp\,\mathrm dx\right)^{\!\ast},
\label{eq:advection-pure-imaginary}
\end{equation}
i.e.\ this integral equals minus its own complex conjugate, so it is pure imaginary and its
real part is exactly zero, for every discrete eigenmode, not merely bounded in sign.
Equation~\eqref{eq:rayleigh-quotient-split-i} then collapses, for every discrete eigenvalue
of the vortical branch and every $\kt\ge0$, to
\begin{equation}
-\,\omega_i\int_{-\infty}^{\infty}\rb|\vperp|^2\,\mathrm dx
=\int_{-\infty}^{\infty}\bar\mu|\D\vperp|^2\,\mathrm dx+\kt^2\!\int_{-\infty}^{\infty}\bar\mu|\vperp|^2\,\mathrm dx\;\ge\;0,
\label{eq:v-no-discrete-instability}
\end{equation}
with equality only for the trivial mode $\vperp\equiv0$. Thus $\omega_i<0$ strictly for every
nontrivial discrete eigenmode of the vortical branch, at every $\kt\ge0$: an unconditional,
self-contained exclusion of discrete instability for this branch, obtained directly from the
governing equation and the base flow's own mass conservation, independently of
\citet{Humpherys2009}, whose spectral-stability result is one-dimensional ($\kt=0$),
and independently of the far-field discriminant maps of \S\ref{sec:results}, which locate
only the continuous spectrum that will be discussed in \S\ref{sec:theory-essential-spectrum}. Given $\omega_i<0$ on
this unconditional footing, dropping the non-negative term $\int\bar\mu|\D\vperp|^2\,\mathrm dx$ in
\eqref{eq:v-no-discrete-instability} gives the sharper, explicitly $\kt$-dependent bound
\begin{equation}
\omega_i\;\le\;-\,\kt^2\,\min_x\frac{\bar\mu}{\rb}\;=\;-\,\frac{\mu_2}{\rho_2}\,\kt^2,
\label{eq:v-kt-bound}
\end{equation}
where the minimum of the kinematic viscosity across the layer is attained at the downstream
state, as verified on the computed profiles at every Mach number from 1.1 to 2 ($\mu_2/\rho_2=0.81,\,0.71,\,0.63$ at
$M_1=1.25,\,1.5,\,2$). Every discrete eigenvalue of the vortical branch therefore lies on or
below the vertex $\omega=-\mathrm i(\mu_2/\rho_2)\kt^2$ of the downstream vortical continuous
spectrum \eqref{eq:vorticity-dispersion}, and this ceiling recedes from the real axis as $\kt^2$.

The bound \eqref{eq:v-kt-bound} belongs to the family of a priori eigenvalue bounds that began with the
classic semicircle theorem of \citet{Howard1961} and were extended to the Orr--Sommerfeld equation by \citet{Joseph1968}.
The analysis method follows the steps of these works in that the eigenfunction is projected on its own equation, the identity is split into real and imaginary parts, and the
confining region follows from the signs of the integrals and is fixed by base-state functionals alone, here $\min_x(\bar\mu/\bar\rho)$ in place of the range of $U$ in Howard's analysis.
As there, the boundary is anchored to the continuous spectrum, here the vertex of the downstream vortical continuous spectrum, and the covariance theorem
plays the role of Squire's transformation in extending the bound to oblique waves.
The differences with Howard's theorem arise from its being inviscid and constraining unstable eigenvalues only, whereas \eqref{eq:v-kt-bound} is viscous, excludes instability
altogether, recedes from the real axis as $\kt^2$ and gives no bound on $\omega_r$. The present confinement is also dissipative, with the advection integral being purely imaginary for a discrete
eigenmode because $\D(\bar\rho\bar u)=0$, while the real part reduces to two non-negative viscous integrals. Finally, the boundary of \eqref{eq:v-kt-bound} is attained by the
continuous spectrum, so the ceiling cannot be lowered for the spectrum as a whole. The relation to the theorems of Howard and Joseph is therefore one of methodological approach, and
\eqref{eq:v-kt-bound} is not a corollary of either.

Within the acoustic--entropy block the same mechanism recurs, but it is not confined to the
$\vpar$-row, since $\kt^2$ enters as an explicit, non-negative reaction term in three of the four
rows of \eqref{eq:gs-at} and only the continuity row carries none. A row-by-row extension of the Rayleigh-quotient argument
does not close for this coupled system, since $\uh$ and $\vpar$
are linked by an off-diagonal shear term linear in $\kt$ whose sign is not determined a
priori. The next section \S\ref{sec:theory-energy-ae} closes the result with a Chu-type
disturbance-energy functional. What is exact is the transformation itself: a three-dimensional disturbance at
$(k_y,k_z)$ is the two-dimensional disturbance at $\kt=(k_y^2+k_z^2)^{1/2}$, the Reynolds-free
counterpart of Squire's transformation, with $\kt$ in the role played by $\Rey$ in the
boundary layer. That the least-damped point of the oblique disturbance's spectrum is never less damped than that of its two-dimensional
counterpart at the same $k_y$ is exact for the vortical branch, whose least-damped point is the vertex of the downstream continuous spectrum
\eqref{eq:vorticity-dispersion}, since every discrete eigenvalue lies below it by \eqref{eq:v-kt-bound}. For the
acoustic--entropy branch the same argument rests on the analysis presented in the next section, \S\ref{sec:theory-energy-ae}, and the numerical verifications
discussed in \S\ref{sec:results-kt-sweep}.

\subsection{A coupled disturbance-energy identity for the acoustic--entropy branch}
\label{sec:theory-energy-ae}

The acoustic--entropy
branch couples four fields, $(\rh,\uh,\vpar,\Th)$, through \eqref{eq:gs-at}, with $\vh\equiv\vpar$, $k_z=0$, $k_y=\kt$.
A coupled identity follows from the disturbance-energy method of \citet{Chu1965}. Each equation of
\eqref{eq:gs-at} is multiplied by the complex conjugate of its own perturbation variable, with weights
\begin{equation}
\beta_\rho=\frac{T_\infty}{\gamma M_1^2\rho_\infty},\qquad
\beta_T=\frac{1}{\gamma(\gamma-1)M_1^2T_\infty},
\label{eq:chu-weights}
\end{equation}
on the continuity and energy rows and unity on the momentum rows, integrated over $x$ and summed. The weights are fixed, uniquely up to a constant, by requiring
the pressure-work exchanges between continuity and $x$-momentum, and between $x$-momentum and energy, to
cancel after integration by parts. The eigenmodes of the constant-coefficient system \eqref{eq:gs-at} are Fourier modes
$\hat{\mathbf q}\propto e^{\mathrm i\xi x}$, which trace the
acoustic--entropy continuous spectrum (\S\ref{sec:theory-essential-spectrum}), so every integral is an average over
one wavelength and no boundary terms arise. The right-hand side of the sum is $\mathrm i\omega$ times twice the disturbance
energy
\begin{equation}
2E\equiv \beta_\rho|\rh|^2+\rho_\infty\big(|\uh|^2+|\vpar|^2\big)+\rho_\infty \beta_T|\Th|^2,
\label{eq:chu-energy}
\end{equation}
a positive-definite form playing the role that $\rho|\vperp|^2$ plays for the vortical branch.
On the left-hand side, every advective self-coupling term and every pressure-work or $\mathrm
ik_y$-weighted coupling between a pair of fields is purely imaginary after integration by
parts. One coupling which does not follow the above pattern is the shear/dilatation term linking
$\uh$ and $\vpar$, weighted by $\lambda_\infty+\tfrac13\mu_\infty$ in both momentum rows of \eqref{eq:gs-at}. With $Z\equiv\int_{-\infty}^{\infty}\hat
u^\ast\D\vpar\,\mathrm dx$ and $\lambda_\infty=0$, this pair sums to the \emph{real}
quantity $\tfrac23\mu_\infty k_y\,\mathrm{Im}(Z)$, of indefinite sign. Equating
real and imaginary parts as in
\eqref{eq:rayleigh-quotient-split} exactly results in the coupled energy balance:
\begin{equation}
-\,\omega_i\cdot2E=\tfrac43\mu_\infty\!\int|\D\uh|^2\mathrm dx+\mu_\infty\!\int|\D\vpar|^2\mathrm dx\notag
\end{equation}
\begin{equation}
\qquad+\frac{\mu_\infty}{(\gamma-1)\Pran M_1^2T_\infty}\!\int|\D\Th|^2\mathrm dx
+\mu_\infty\kt^2\!\int|\uh|^2\mathrm dx\notag
\end{equation}
\begin{equation}
\qquad+\tfrac43\mu_\infty\kt^2\!\int|\vpar|^2\mathrm dx
+\frac{\mu_\infty}{(\gamma-1)\Pran M_1^2T_\infty}\kt^2\!\int|\Th|^2\mathrm dx
\;+\;\tfrac23\mu_\infty\kt\,\mathrm{Im}(Z),
\label{eq:chu-identity}
\end{equation}
at every point of the acoustic--entropy continuous spectrum. Every term on the right of
\eqref{eq:chu-identity} is non-negative except the final shear-coupling term, which is controlled by two of the others, as follows.
By the Cauchy--Schwarz inequality,
$|\mathrm{Im}(Z)|\le\|\uh\|_{L^2}\|\D\vpar\|_{L^2}$, and by Young's inequality with
parameter $\kt$,
\begin{equation}
\|\uh\|_{L^2}\|\D\vpar\|_{L^2}=\big(\kt^{1/2}\|\uh\|_{L^2}\big)\big(\kt^{-1/2}\|\D\vpar\|_{L^2}\big)
\le\tfrac12\Big(\kt\|\uh\|_{L^2}^2+\kt^{-1}\|\D\vpar\|_{L^2}^2\Big),
\label{eq:young-bound}
\end{equation}
so that
\begin{equation}
\Big|\tfrac23\mu_\infty\kt\,\mathrm{Im}(Z)\Big|\le\tfrac13\mu_\infty\kt^2\!\int|\uh|^2\mathrm dx
+\tfrac13\mu_\infty\!\int|\D\vpar|^2\mathrm dx.
\label{eq:cross-bound}
\end{equation}
Substituting \eqref{eq:cross-bound} into \eqref{eq:chu-identity}
results in
\begin{equation}
-\,\omega_i\cdot2E\;\ge\;\tfrac43\mu_\infty\!\int|\D\uh|^2\mathrm dx+\tfrac23\mu_\infty\!\int|\D\vpar|^2\mathrm dx\notag
\end{equation}
\begin{equation}
\qquad+\frac{\mu_\infty}{(\gamma-1)\Pran M_1^2T_\infty}\!\int\big(|\D\Th|^2+\kt^2|\Th|^2\big)\mathrm dx\notag
\end{equation}
\begin{equation}
\qquad\qquad+\;\tfrac23\mu_\infty\kt^2\!\int|\uh|^2\mathrm dx+\tfrac43\mu_\infty\kt^2\!\int|\vpar|^2\mathrm dx\;\ge\;0,
\label{eq:chu-result}
\end{equation}
for every far-field eigenmode and every $\kt$. Every term on the right is non-negative. The right-hand side vanishes only for
$\uh=\vpar=\Th=0$, a pure density perturbation, which the momentum rows admit only at $\xi=\kt=0$. Hence $\omega_i<0$ strictly at every
point of the acoustic--entropy continuous spectrum other than the origin, at both far-field
states and for every $\kt$. The bound does not force damping to grow like $\kt^2$, since $\beta_\rho|\rh|^2$ in $2E$ has no
counterpart on the right, so a mode carried mostly by $\rh$ can stay weakly damped at large $\kt$, which is the
saturation found in \S\ref{sec:results-kt-sweep}.

Two limits apply to this result. First, it concerns the continuous spectrum only: for discrete eigenvalues of the full, $x$-varying operator the balance acquires
production terms proportional to $\D\ub$, $\D\Tb$, of indefinite sign. Absence of unstable discrete
eigenvalues is established at $\kt=0$ by \citet{Humpherys2009} through asymptotic estimates and numerical Evans-function computations, and for $\kt>0$ is supported only numerically, as seen in 
appendix~\ref{app:evp-validation} and \S\ref{sec:transient-results}. Second, the bound holds for every mode 
but does not make $\omega_i$ monotone in $\kt$, since the
eigenfunction and $2E$ also vary with $\kt$, so the ordering in $\kt$ rests on the bound and on the dense numerical sweep of \S\ref{sec:results-kt-sweep}.

\subsection{Companion-matrix discriminant}
\label{sec:theory-companion-discriminant}\label{sec:theory-fardiag}

Casting \eqref{eq:gs-vorticity} and \eqref{eq:gs-at} in first-order companion form gives,
at every complex $\omega$ and real $k_y$, and at each of the two uniform far-field states
$(\rho_\infty,u_\infty,T_\infty,\mu_\infty)|_{\rm cold,hot}$ related by the Rankine--Hugoniot
conditions \eqref{eq:RH} and
$\rho_\infty=1/u_\infty$, two constant matrices, the $7\times7$ matrix $\mathbf M_{\rm AE}(\omega,k_y)$
and the $2\times2$ matrix $\mathbf M_V(\omega,k_y)$, such that, for the acoustic--entropy
state vector $\hat{\mathbf q}_{\rm AE}=(\rh,\uh,\D\uh,\vh,\D\vh,\Th,\D\Th)^{\rm T}$ and the
vortical state vector $\hat{\mathbf q}_V=(\wh,\D\wh)^{\rm T}$ of \S\ref{sec:theory-gs}, the
far-field companion system and its eigenmodes take the form
\begin{subequations}
\label{eq:companion-ansatz}
\begin{align}
\begin{aligned}
\D\hat{\mathbf q}_{\rm AE}&=\mathbf M_{\rm AE}(\omega,k_y)\,\hat{\mathbf q}_{\rm AE},\\
\hat{\mathbf q}_{\rm AE}(x)&=\hat{\mathbf q}_{{\rm AE},j}\,e^{\sigma_jx},\\
\mathbf M_{\rm AE}(\omega,k_y)\,\hat{\mathbf q}_{{\rm AE},j}&=\sigma_j\,\hat{\mathbf q}_{{\rm AE},j},
\end{aligned}
\label{eq:companion-ansatz-ae}\\[8pt]
\begin{aligned}
\D\hat{\mathbf q}_V&=\mathbf M_V(\omega,k_y)\,\hat{\mathbf q}_V,\\
\hat{\mathbf q}_V(x)&=\hat{\mathbf q}_{V,j}\,e^{\sigma_jx},\\
\mathbf M_V(\omega,k_y)\,\hat{\mathbf q}_{V,j}&=\sigma_j\,\hat{\mathbf q}_{V,j},
\end{aligned}
\label{eq:companion-ansatz-v}
\end{align}
\end{subequations}
so that, for $\bullet={\rm AE},V$, the eigenvalues $\sigma_j$ of $\mathbf M_\bullet$ are the spatial decay or growth
rates of every admissible far-field eigenmode. This is the mechanism identified by
\citet{GroschSalwen1978} in the free-stream of the incompressible flat-plate boundary layer, where
the coefficients of the linear operator tend to constant
far-field values and its far-field eigenmodes are exactly those of the resulting algebraic
companion matrix, and it is this matrix, not the variable-coefficient operator itself,
that organizes the continuous spectrum and any branch points or embedded discrete
eigenvalues on a semi-infinite or infinite domain. Adapting the discriminant diagnostic
introduced for the outer problem in \S5 of \citet{Theofilis2026} to the present,
$\Rey$-free equations, the complex $\omega$-plane is scanned on a uniform grid of
$(\omega_r,\omega_i)$ values,
\begin{equation}
d_\bullet(\omega;k_y)=\min\Big(\min_j\big|\mathrm{Re}\,\sigma_j^{\rm cold}\big|,\
\min_j\big|\mathrm{Re}\,\sigma_j^{\rm hot}\big|\Big),\qquad \bullet={\rm AE},V,
\label{eq:discriminant}
\end{equation}
the minimum, over both far-field states and every companion eigenvalue, of the distance of
$\sigma_j$ from the imaginary axis: near-zero valleys of $d_\bullet$ trace boundaries of
the continuous spectrum and mark candidates for branch points or embedded discrete
eigenvalues of each branch. By the covariance theorem of \S\ref{sec:theory-theorem}, both branches may be evaluated at
$k_z=0,\,k_y=k_t$ without loss of generality.

Both companion matrices are, by construction, affine in $\omega$ with a purely
imaginary slope, $\mathbf M_\bullet(\omega,k_y)=\mathbf A_\bullet(k_y)+\mathrm i\omega\,\mathbf B_\bullet(k_y)$
with $\mathbf A_\bullet,\mathbf B_\bullet$ real, after the similarity transformation $\vh=\mathrm i\tilde v$, which removes the explicit factors $\mathrm ik_y$ without changing the eigenvalues, and since $\omega$ enters \eqref{eq:gs-vorticity}
and \eqref{eq:gs-at} only through the Fourier convention $\partial_t\to-\mathrm i\omega$ acting
on real base-flow coefficients. Consequently, at every fixed $\omega_i$,
$\mathbf M_\bullet(-\omega_r+\mathrm i\omega_i,k_y)=\overline{\mathbf M_\bullet(\omega_r+\mathrm i\omega_i,k_y)}$,
so the companion eigenvalues at $-\omega_r+\mathrm i\omega_i$ are the complex conjugates of those at $\omega_r+\mathrm i\omega_i$,
$\sigma_j(-\omega_r+\mathrm i\omega_i)=\overline{\sigma_j(\omega_r+\mathrm i\omega_i)}$, and since
conjugation leaves the real part unchanged,
\begin{equation}
d_\bullet(-\omega_r+\mathrm i\omega_i;k_y)=d_\bullet(\omega_r+\mathrm i\omega_i;k_y)
\qquad\text{exactly, for }\bullet={\rm AE},V,
\label{eq:discriminant-mirror}
\end{equation}
at every $k_y$. Every discriminant map below is therefore mirror-symmetric about the
imaginary $\omega$-axis by construction, independently of the base-flow or Mach number: a
routine structural check, and one that halves the computational domain needed for any
such map to $\omega_r\ge0$.

\subsection{Exact far-field dispersion relations}
\label{sec:analytical-dispersion}\label{sec:theory-dispersion-v}\label{sec:theory-dispersion-ae}

An alternative to the discriminant scan of \S\ref{sec:theory-companion-discriminant} for locating the continuous spectrum is discussed in this
section. It rests on the fact that 
the far-field equation \eqref{eq:gs-vorticity} is a constant-coefficient ODE in
$\wh(x)$, and its eigenmodes may be sought analytically following two complementary routes. First, $\omega$ and $k_y$ are fixed and
a solution is sought for the spatial rate $\sigma$ of $\wh\propto e^{\sigma x}$ from the quadratic
\begin{equation}
\mu_\infty\sigma^2-\rho_\infty u_\infty\sigma+\big(\mathrm i\omega\rho_\infty-\mu_\infty k_y^2\big)=0,
\label{eq:vorticity-sigma-quadratic}
\end{equation}
whose roots are the eigenvalues of the companion matrix
$\mathbf M_V(\omega,k_y)$ of \eqref{eq:companion-ansatz-v}. This route admits any complex $\sigma$ and is the one
needed for isolated discrete eigenvalues and for the discriminant $d_V$ of \eqref{eq:discriminant}.
The second route fixes $k_y$ and asks which \emph{real} frequencies
$\omega$ admit a bounded, oscillatory mode, $\wh=e^{\mathrm i\xi x}$ with $\xi\in\mathbb{R}$, the defining signature of the
continuous spectrum. This is $\sigma=\mathrm i\xi$ in \eqref{eq:vorticity-sigma-quadratic}, and
solving for $\omega$ gives the exact, closed-form dispersion relation
\begin{equation}
\omega(\xi,k_y)=u_\infty\xi-\mathrm i\,\frac{\mu_\infty}{\rho_\infty}\big(\xi^2+k_y^2\big),
\qquad\xi\in\mathbb{R},
\label{eq:vorticity-dispersion}
\end{equation}
at each of the two far-field states $(\rho_\infty,u_\infty,\mu_\infty)|_{\rm
cold,hot}$ of \S\ref{sec:theory-companion-discriminant}. It describes a transverse shear
wave advected at $u_\infty$ and damped at the rate $(\mu_\infty/\rho_\infty)(\xi^2+k_y^2)$, strictly increasing in
$\xi^2$ and in $k_t^2=k_y^2$. Every point of the vortical continuous spectrum is therefore damped, increasingly so with
transverse wavenumber, recovering without any
discriminant search the $k_t$-monotonicity that \S\ref{sec:theory-theorem-consequence}
establishes by a Rayleigh-quotient argument. The two results are complementary: the latter bounds
the decay rate of $L^2$ eigenmodes, while \eqref{eq:vorticity-dispersion} requires no decay at all and locates
the extended continuous-spectrum curve itself \citep{Humpherys2009}.
Because $\mu_\infty/\rho_\infty$ is constant in the far field, $\kt$ shifts every
point of the continuous spectrum rigidly, by $-\mathrm i(\mu_\infty/\rho_\infty)\kt^2$ at either far-field state.
This contrasts with the discrete eigenvalues, for which $\bar\mu/\rb$ varies across the layer
(\S\ref{sec:theory-theorem-consequence}).

For the acoustic--entropy branch, writing
$(\rh,\uh,\vpar,\Th)\propto e^{\mathrm i\xi x}$, $\xi\in\mathbb{R}$, and
$\Delta\equiv u_\infty\xi-\omega$ in \eqref{eq:gs-at}, with $\lambda_\infty=0$
(\S\ref{sec:theory-baseflow}), gives the linear algebraic
system
\begin{equation}
\mathbf A(\xi,\omega;k_y)\,\hat{\mathbf q}=\mathbf 0,\qquad \hat{\mathbf q}\equiv(\rh,\uh,\vpar,\Th)^{\rm T},
\label{eq:ae-dispersion-matrix}
\end{equation}
where the explicit $4\times4$ matrix is
\begin{equation}
\resizebox{0.95\linewidth}{!}{$\displaystyle
\mathbf A=\begin{pmatrix}
\Delta & \rho_\infty\xi & k_y\rho_\infty & 0\\[4pt]
\dfrac{T_\infty}{\gamma M_1^2}\mathrm i\xi &
\mathrm i\rho_\infty\Delta+\mu_\infty\big(\tfrac43\xi^2+k_y^2\big) &
\tfrac13\mu_\infty k_y\xi &
\dfrac{\rho_\infty}{\gamma M_1^2}\mathrm i\xi\\[8pt]
\mathrm ik_y\dfrac{T_\infty}{\gamma M_1^2} &
\tfrac13\mu_\infty k_y\xi &
\mathrm i\rho_\infty\Delta+\mu_\infty\big(\xi^2+\tfrac43k_y^2\big) &
\mathrm ik_y\dfrac{\rho_\infty}{\gamma M_1^2}\\[8pt]
0 & \mathrm i(\gamma-1)\rho_\infty T_\infty\xi & \mathrm ik_y(\gamma-1)\rho_\infty T_\infty &
\mathrm i\rho_\infty\Delta+\dfrac{\gamma\mu_\infty}{\Pran}\big(\xi^2+k_y^2\big)
\end{pmatrix}$}
.
\label{eq:ae-dispersion-matrix-explicit}
\end{equation}
A nontrivial mode exists at real $\xi$ only where
\begin{equation}
\det\mathbf A(\xi,\omega;k_y)=0,
\label{eq:ae-dispersion-det}
\end{equation}
the exact dispersion relation of the acoustic--entropy branch. Since $\omega$ enters $\mathbf A$ only through $\Delta$, which appears once in each of its four diagonal entries, for every real $\xi$,
$\det\mathbf A$ is a quartic polynomial in $\Delta$, and therefore in $\omega$, as $\Delta$ is linear in $\omega$. Although no elementary closed form of the quartic exists, this equation can be solved
robustly by writing
$\mathbf A=\mathbf A_0(\xi;k_y)-\omega\,\mathbf C$, with $\mathbf A_0$ equal to
$\mathbf A$ at $\Delta\to u_\infty\xi$ and
$\mathbf C=\mathrm{diag}(1,\mathrm i\rho_\infty,\mathrm i\rho_\infty,\mathrm i\rho_\infty)$
constant and invertible, so that \eqref{eq:ae-dispersion-det} is the
generalized eigenvalue problem $\mathbf A_0(\xi;k_y)\hat{\mathbf q}=\omega\mathbf C\hat{\mathbf q}$, whose
four eigenvalues $\omega_j(\xi,k_y)$ give the continuous spectrum exactly.
At $k_y=0$ the $\vpar$-row decouples and returns the vortical form
$\omega=u_\infty\xi-\mathrm i(\mu_\infty/\rho_\infty)\xi^2$, while $(\rh,\uh,\Th)$
remain coupled into a cubic. For $k_y\neq0$ all four fields are coupled, so the $k_t$-dependence of the roots cannot be read off by
inspection, and \S\ref{sec:theory-energy-ae} supplies the damping result instead.

\subsection{The continuous spectrum and the translational neutral mode}
\label{sec:theory-essential-spectrum}\label{sec:theory-essential-spectrum-operator}\label{sec:theory-essential-spectrum-essential}\label{sec:theory-essential-spectrum-translation}\label{sec:theory-essential-spectrum-fredholm}

As noted in \S\ref{sec:theory-lst}, equation \eqref{eq:lnse} can take the modal eigenvalue problem form $\mathbf L(k_y,k_z)\,\hat{\mathbf q}=\mathrm i\omega\,\mathbf B\,\hat{\mathbf q}$ with
$\mathbf B=\mathrm{diag}(1,\rb,\rb,\rb,\rb)$. 
The rotation \eqref{eq:rotation-inner}, an orthogonal matrix $\mathbf R(\theta)$ commuting with $\mathbf B$,
block-diagonalizes $\mathbf L$ via the covariance theorem,
\begin{equation}
\mathbf R(\theta)^{\rm T}\,\mathbf L(k_y,k_z)\,\mathbf R(\theta)\;=\;\mathbf L_{\rm AE}(\kt)\,\oplus\,\mathbf L_V(\kt),
\qquad \mathbf B=\mathbf B_{\rm AE}\oplus \mathbf B_V,
\label{eq:operator-identity}
\end{equation}
for every $(k_y,k_z)\neq(0,0)$, where $\mathbf L_{\rm AE}$ acts on $(\rh,\uh,\vpar,\Th)^{\rm T}$ and $\mathbf L_V\equiv\mathcal L$ of
\eqref{eq:operator-def} acts on $\vperp$. 
Here $\oplus$ denotes the direct sum, that is, the block-diagonal operator in which $\mathbf L_{\rm AE}$ acts on the in-plane components
$(\rh,\uh,\vpar,\Th)^{\rm T}$ and $\mathbf L_V$ on the out-of-plane component $\vperp$, with no coupling between the two sets, and likewise for $\mathbf B_{\rm AE}\oplus\mathbf B_V$.
Consequently,
\begin{equation}
\mathrm{spec}\,\big(\mathbf L(k_y,k_z),\mathbf B\big)=\mathrm{spec}\,\big(\mathbf L_{\rm AE}(\kt),\mathbf B_{\rm AE}\big)
\,\cup\,\mathrm{spec}\,\big(\mathbf L_V(\kt),\mathbf B_V\big).
\label{eq:spectrum-union}
\end{equation}
At a value of $\omega$ shared by both branches the eigenspace is the direct sum of the two invariant subspaces, so an eigenvector may combine the two kinds, but each component is an eigenvector of one branch. At $\kt=0$ the rotation is not needed, since $\vh$ and $\wh$ each satisfy the vortical equation and $(\rh,\uh,\Th)$ form the
acoustic--entropy block.

Both operators have coefficients that tend to those of the two uniform far-field states as $x\to\mp\infty$, so
their continuous spectrum, denoted $\sigma_{\rm ess}$ (the essential spectrum of the mathematical literature), is fixed by these two constant-coefficient limits, as in
the Evans-function theory of viscous shock profiles \citep{ZumbrunSerre1999,Humpherys2009}, according to which $\omega$ lies in it if and only if a limit admits a bounded,
non-decaying mode $\propto e^{\mathrm i\xi x}$, $\xi\in\mathbb R$. This is the construction of
\S\ref{sec:theory-dispersion-v}, so that
\begin{equation}
\sigma_{\rm ess}\big(\mathbf L_V(\kt),\mathbf B_V\big)=
\big\{\omega(\xi,\kt):\xi\in\mathbb R\big\}_{\rm cold}\,\cup\,
\big\{\omega(\xi,\kt):\xi\in\mathbb R\big\}_{\rm hot},
\label{eq:essential-spectrum-v}
\end{equation}
the two branches of \eqref{eq:vorticity-dispersion}, and the acoustic--entropy essential
spectrum is the union over both states of the four root loci of
\eqref{eq:ae-dispersion-det}. These loci are what the term ``continuous spectrum'' denotes throughout, so that in this problem the two terms name the same set.

For the one-dimensional ($\kt=0$) shock profile no
unstable discrete eigenvalue ($\omega_i>0$) exists \citep{Humpherys2009}. For $\kt>0$, \eqref{eq:v-no-discrete-instability} gives $\omega_i<0$ for
every discrete vortical eigenmode, and \eqref{eq:v-kt-bound} places it
below the vertex of the downstream vortical continuous spectrum. For the acoustic--entropy
branch, \S\ref{sec:theory-energy-ae} proves that the continuous spectrum lies in
$\omega_i<0$ away from the origin, and no unstable discrete
eigenvalue of the full operator is found in the computations of
appendix~\ref{app:evp-validation} and \S\ref{sec:transient-results}, although none is excluded analytically.
Subject to that evidence, no three-dimensional eigenmode is
unstable, and the discriminant maps of \S\ref{sec:results} describe the least-damped part of
the spectrum for $\kt>0$.

The steady equations \eqref{eq:baseflow-inner} are autonomous in $x$, so $\bar{\mathbf q}(x-x_0)$ is a solution for every $x_0$, and
differentiating with respect to $x_0$ places $\bar{\mathbf q}'(x)$ in the kernel of the linearization \eqref{eq:lnse} at
$\omega=k_y=k_z=0$:
\begin{equation}
\mathbf L(0,0)\,\hat{\mathbf q}_{\rm transl}=0,\qquad
\hat{\mathbf q}_{\rm transl}\equiv\big(\D\rb,\D\ub,0,0,\D\Tb\big)^{\rm T}.
\label{eq:translation-mode}
\end{equation}
This zero mode, generic to translation-invariant fronts \citep{Humpherys2009}, has $\vh=\wh=0$ and belongs to the
acoustic--entropy branch. It decays as $x\to\pm\infty$ and is unrelated to the far-field
degeneracy $\mathbf A_0(0,0)=\mathbf 0$ of equation \eqref{eq:ae-origin-zero} in appendix~\ref{app:origin-boundedness}, but the two meet at the same point, since
$\omega=0$ lies on the continuous spectrum of both branches at $\kt=0$. The mode is therefore an
eigenvalue embedded where the continuous spectrum touches the real axis. It reflects only the translation freedom of the layer and plays no role for $\kt>0$.

For $\omega\notin\sigma_{\rm ess}$, the operator
$\mathbf L_\bullet(\kt)-\mathrm i\omega \mathbf B_\bullet$ on $L^2(\mathbb R;\mathbf B_\bullet\,\mathrm dx)$ is Fredholm of index zero
\citep{ZumbrunSerre1999,Humpherys2009}, since the constant coefficients at each end fix a
hyperbolic splitting of complementary
dimension in the component of the complement of $\sigma_{\rm ess}$ that contains the half-plane $\omega_i>0$. The index can change only where this
dichotomy breaks down, i.e.\ where a limiting companion matrix has a purely imaginary
eigenvalue, which by \eqref{eq:discriminant} is the zero set of $d_\bullet$. A discrete eigenvalue would instead be an isolated zero of the Evans
function, a property of the whole layer, so $d_\bullet$ alone cannot exclude one.

\begin{figure}
\centering
\includegraphics[width=0.62\textwidth]{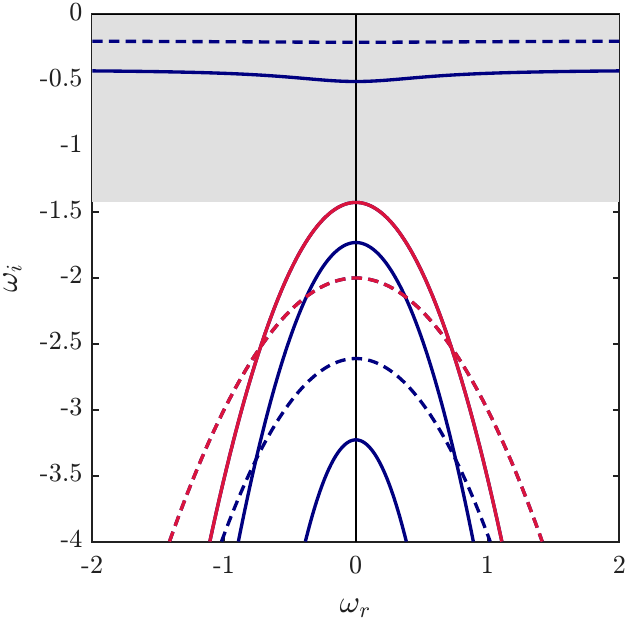}
\caption{Complex $\omega$-plane at $(M_1,\kt)=(1.5,\sqrt2)$. 
Continuous spectra of the vortical (red) and acoustic--entropy (blue) branches, downstream (solid) and upstream (dashed). Gray denotes the region free of discrete vortical eigenvalues.}
\label{fig:vortical-ceiling}
\end{figure}

This section closes with figure~\ref{fig:vortical-ceiling}, which gathers its analytical results in the complex
$\omega$-plane for the monatomic gas of \S\ref{sec:theory-baseflow} at parameters $\gamma=5/3$, $\Pran=2/3$, $M_1=1.5$ and $\kt=\sqrt2$. By the covariance theorem the picture holds for every pair $(k_y,k_z)$ with
$k_y^2+k_z^2=\kt^2=2$, and its two families of curves are the two branches into which the theorem splits every eigenmode.
The vortical continuous spectrum lines, the parabolas of \eqref{eq:vorticity-dispersion}, are lifted rigidly by $\kt^2$, and the vertex of the
downstream one sets the ceiling \eqref{eq:v-kt-bound}, which appears in the figure as the lower edge of the gray region.
At these parameters every discrete vortical eigenvalue lies at $\omega_i\le-(\mu_2/\rho_2)\kt^2=-1.43$, so the gray region,
which contains the whole real axis, holds none.
The acoustic--entropy continuous spectrum, the root loci of \eqref{eq:ae-dispersion-det}, lies in $\omega_i<0$ by
\eqref{eq:chu-result} but does not recede with $\kt$. Its least-damped point sits at the plateau \eqref{eq:ae-plateau} that will be discussed in \S\ref{sec:results-kt-sweep}. At $M_1=1.5$ this plateau lies at
$\omega_i=-0.200$ upstream and $-0.419$ downstream. At $\kt=\sqrt2$ the vortical branch has therefore
retreated below the acoustic--entropy plateau, so the least-damped part of the spectrum belongs to the acoustic--entropy
branch.

\section{Modal results}
\label{sec:results}

This section presents the far-field continuous spectra predicted by the covariance theorem,
at the wavenumber pair $k_z=0$, $k_y=\kt$ throughout, which by the theorem represents every pair $(k_y,k_z)$ with the same $\kt$.
Section~\ref{sec:modal-validation} verifies the Reynolds-free and covariance results against the rescaled
outer problem and an unrotated computation, while appendix~\ref{app:dispersion-check} verifies the numerically located continuous spectra against the exact
dispersion relations of \S\ref{sec:analytical-dispersion}.
Section~\ref{sec:results-mach} discusses the Mach-number dependence of spectra in the limited range of physical validity of the
weak shock layer model, and Sections~\ref{sec:results-squire}--\ref{sec:results-kt-sweep} examine the dependence of spectra on the
transverse wavenumber.
All results of this section are semi-analytic. The continuous spectra are located by evaluating, at each point of a grid in the complex
$\omega$-plane, the eigenvalues of the far-field companion matrices of \S\ref{sec:theory-companion-discriminant}, so that no discretization
of the eigenvalue problem over the shock layer is involved. Computing the spectrum instead by spectral collocation of the full,
$x$-dependent operators on a truncated domain is a different and much less tractable task. The continuous spectrum then appears as a dense
cluster of discrete eigenvalues that reflects the truncation, and convergence in the domain extent is slow, as shown in
appendix~\ref{app:evp-validation}.

\subsection{Numerical verification of the formulation}
\label{sec:modal-validation}

Two numerical tests address the central claims of the theory directly.
The first concerns the elimination of the Reynolds number from the far-field problem discussed in \S\ref{sec:theory-baseflow}.
The discriminants $d_{\rm AE}$ and $d_V$ computed from the Reynolds-free companion matrices are compared, at $\Rey=1$, $100$ and $10^4$,
with those computed from the independently built outer, Reynolds-dependent matrices of \citet{Theofilis2026} after rescaling to inner units.
The second concerns the covariance theorem discussed in \S\ref{sec:theory-theorem}.
The eigenvalues of a $9\times9$ companion matrix, constructed from the linearized equations without rotating the wavenumber vector, are compared with the
union of the eigenvalues of $\mathbf M_{\rm AE}$ and $\mathbf M_V$, for both far-field states, five Mach numbers and randomly split $\kt$.
The first test is a consistency check of the formulation.
The second tests the rotation itself, which the maps of the following subsections assume.

\subsubsection{Reynolds-free versus rescaled outer spectra}
\label{sec:theory-reynolds-collapse}

\begin{figure}
\centering
\includegraphics[width=\textwidth]{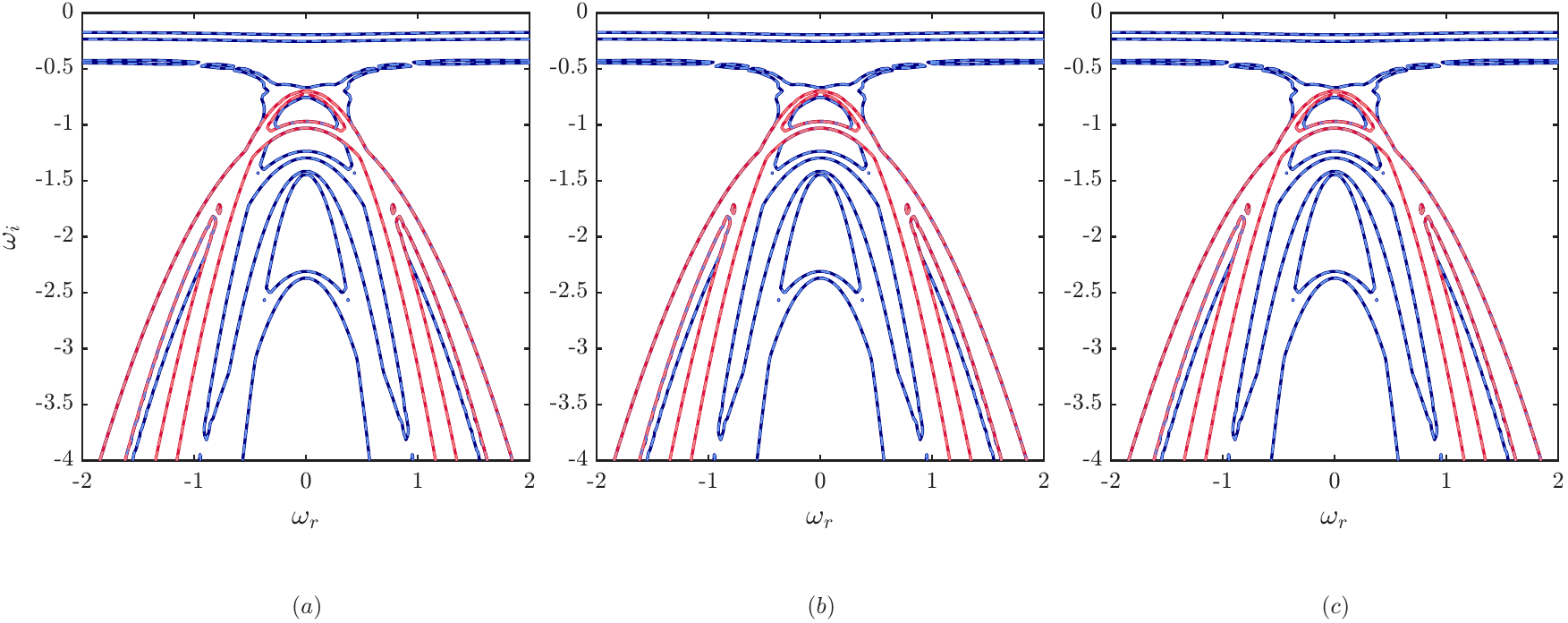}
\caption{Discriminant contours $d_{\rm AE}=0.03$ (blue) and $d_V=0.03$ (red) at
$M_1=1.5,\,k_y=1$: Reynolds-free (solid) versus rescaled
$\Rey$-dependent (dashed) results of \citet{Theofilis2026}, on a $140\times110$ grid in
$(\omega_r,\omega_i)$: (a) $\Rey=1$, (b) $\Rey=100$, (c) $\Rey=10^4$.}
\label{fig:collapse}
\end{figure}

The Reynolds elimination of \S\ref{sec:theory-baseflow} asserts that the outer companion systems, evaluated at any $\Rey$
and rescaled through \eqref{eq:rescaling}, reproduce \eqref{eq:gs-vorticity}--\eqref{eq:gs-at}
identically. For $M_1=1.5$, $k_y=1$ (inner), $d_{\rm AE}$ and
$d_V$ are computed twice at each of $\Rey=1,100,10^4$, once from the Reynolds-free
matrices $\mathbf M_{\rm AE},\mathbf M_V$ of \S\ref{sec:theory-companion-discriminant} directly, and once
from the corresponding outer companion matrices of \citet{Theofilis2026}, built independently from the outer
form of the stability equations, evaluated at $\Om=\Rey\,\omega,\,\Ky=\Rey\,k_y$ and mapped
back to inner units by \eqref{eq:rescaling}. The $d=0.03$ contours of the two branches,
solid for the Reynolds-free map, dashed for the rescaled outer map, are indistinguishable
at every $\Rey$ in figure~\ref{fig:collapse}; the maximum pointwise difference between the
two discriminant fields is exactly zero at $\Rey=1$ (where the rescaling is trivial), and
$1.3\times10^{-14}$ ($d_{\rm AE}$) and $1.4\times10^{-15}$ ($d_V$) at $\Rey=100$, and
$1.5\times10^{-14}$ ($d_{\rm AE}$) and $1.8\times10^{-15}$ ($d_V$) at $\Rey=10^4$, which is
machine precision at every Reynolds number tested, spanning four decades.

This agreement is a consistency check, not an independent physical finding: \eqref{eq:rescaling}
is an exact, invertible change of variables applied term by term to the governing
equations of \citet{Theofilis2026}, so the two companion systems are
\emph{algebraically} identical and any two correct discretizations must agree to machine precision.
Consequently, a single Reynolds-free calculation, at fixed $M_1$, $\gamma$, $\Pran$ and
viscosity law and swept over $k_t$, determines the complete non-dimensional spectral
structure of the shock layer. A dimensional prediction for a specific upstream
state requires only the two conversion factors \eqref{eq:dimensional-conversion}, with no further eigenvalue problem to solve and no reference to an outer length or
Reynolds number at any point.

\subsubsection{Unrotated computation of the decoupling}
\label{sec:results-rotation-validation}

The maps of \S\ref{sec:results} and the dispersion check of appendix~\ref{app:dispersion-check} are computed in the rotated
variables and cannot detect potential errors in the rotation itself. As an
independent check, the far-field limit of \eqref{eq:lnse-cont}--\eqref{eq:lnse-energy} is
written, as in \S\ref{sec:theory-gs} but with $k_z$ retained, as the companion system
\begin{equation}
\D\hat{\mathbf q}_{\rm full}=\mathbf M_{\rm full}(\omega,k_y,k_z)\,\hat{\mathbf q}_{\rm full},\qquad
\hat{\mathbf q}_{\rm full}=(\rh,\uh,\D\uh,\vh,\D\vh,\wh,\D\wh,\Th,\D\Th)^{\rm T},
\label{eq:full-companion}
\end{equation}
i.e. a $9\times9$ matrix, built directly from the five equations without the rotation, that reduces symbolically to
$\mathbf M_{\rm AE}\oplus\mathbf M_V$ at $k_z=0$. For $k_z\neq0$ and any trial $\omega$, the covariance theorem
requires the nine eigenvalues of $\mathbf M_{\rm full}$ to coincide with the union of the seven eigenvalues of
$\mathbf M_{\rm AE}(\omega,\kt)$ and the two of $\mathbf M_V(\omega,\kt)$, however $\kt$ is split between $k_y$ and $k_z$.
Table~\ref{tab:rotation-validation} tests this for $M_1=1.5$, the downstream state, a trial $\omega=0.3-0.2\mathrm i$ that
lies on no dispersion curve, and $\kt=\sqrt2$, split in two ways, $(k_y,k_z)=(1,1)$ and $(\sqrt2,0)$. The eigenvalues of
$\mathbf M_{\rm full}$ for the two splits and those of the rotated $7\oplus2$ union agree to the eight digits shown,
including two double eigenvalue pairs.

\begin{table}
\centering
\caption{Eigenvalues of $\mathbf M_{\rm full}$ \eqref{eq:full-companion} at two $(k_y,k_z)$ splits of
$\kt=\sqrt2$, and of the rotated $\mathbf M_{\rm AE}\oplus\mathbf M_V$; $M_1=1.5$,
downstream state, $\omega=0.3-0.2\mathrm i$.}
\label{tab:rotation-validation}
\begin{tabular}{cccc}
\toprule
& $(k_y,k_z)=(1,1)$ & $(k_y,k_z)=(\sqrt2,0)$ & $\mathbf M_{\rm AE}\oplus\mathbf M_V$ \\
\midrule
1 & $-1.14965478+0.06803706\mathrm i$ & $-1.14965478+0.06803706\mathrm i$ & $-1.14965478+0.06803706\mathrm i$\\
2 & $-0.97306081+0.15218431\mathrm i$ & $-0.97306081+0.15218431\mathrm i$ & $-0.97306081+0.15218431\mathrm i$\\
3 & $-0.97306081+0.15218431\mathrm i$ & $-0.97306081+0.15218431\mathrm i$ & $-0.97306081+0.15218431\mathrm i$\\
4 & $-0.95000323+0.04886266\mathrm i$ & $-0.95000323+0.04886266\mathrm i$ & $-0.95000323+0.04886266\mathrm i$\\
5 & $-0.49862412+0.56545451\mathrm i$ & $-0.49862412+0.56545451\mathrm i$ & $-0.49862412+0.56545451\mathrm i$\\
6 & $\phantom{-}1.49877263-0.05638874\mathrm i$ & $\phantom{-}1.49877263-0.05638874\mathrm i$ & $\phantom{-}1.49877263-0.05638874\mathrm i$\\
7 & $\phantom{-}1.66437291-0.11167976\mathrm i$ & $\phantom{-}1.66437291-0.11167976\mathrm i$ & $\phantom{-}1.66437291-0.11167976\mathrm i$\\
8 & $\phantom{-}1.79097862-0.15218431\mathrm i$ & $\phantom{-}1.79097862-0.15218431\mathrm i$ & $\phantom{-}1.79097862-0.15218431\mathrm i$\\
9 & $\phantom{-}1.79097862-0.15218431\mathrm i$ & $\phantom{-}1.79097862-0.15218431\mathrm i$ & $\phantom{-}1.79097862-0.15218431\mathrm i$\\
\bottomrule
\end{tabular}
\end{table}

The comparison was repeated over a randomized sweep of 5 Mach numbers ($M_1=1.1$--$2.0$), both far-field states,
and 8 random triples $(\kt,\theta,\omega)$ at each, with $\kt\in(0.2,2.5)$, direction
$\theta\in(0,\tfrac\pi2)$ splitting $\kt$ into $(k_y,k_z)=\kt(\cos\theta,\sin\theta)$, and a
trial $\omega$ spanning both stable and unstable
half-planes. In all 80 trials the 720 eigenvalue pairs, matched by nearest neighbor, agree to within
$1.9\times10^{-14}$, at the level of round-off. Since $\mathbf M_{\rm full}$ is built without the
rotation, this confirms the covariance theorem independently.

\subsection{Mach-number dependence of the spectrum}
\label{sec:results-mach}

Results in this and the next subsection have been obtained on a uniform grid 
$\omega_r\in[-4,4],\,\omega_i\in[-4,0]$, with $k_y=\kt=1$ in figure~\ref{fig:mach} and $k_y=\kt=1,\sqrt2,2$ in figure~\ref{fig:squire},
and the monatomic-gas closure of \S\ref{sec:theory-baseflow}, and use the companion-matrix diagnostics
$\mathbf M_{\rm AE},\mathbf M_V,d_{\rm AE},d_V$ of \S\ref{sec:theory-companion-discriminant},
verified in \S\ref{sec:modal-validation} and appendix~\ref{app:dispersion-check}.
The computations extend beyond what is shown, since the figures display only the window $|\omega_r|\le2$ of the computed range.
The contours are drawn at the level $d_\bullet=0.03$ rather than at $d_\bullet=0$, which a grid scan cannot resolve.
Each therefore lies slightly off the continuous spectrum of \S\ref{sec:theory-essential-spectrum}, the zero set of $d_\bullet$,
by an offset quantified in \S\ref{sec:results-kt-sweep}.
Grid convergence of the results presented has been ensured. Refining the $220\times180$ grid
of figures~\ref{fig:mach} and \ref{fig:squire} to $440\times360$ and $880\times720$, for all five
$(M_1,\kt)$ pairs shown, moves the uppermost point of every $d_{\rm AE}=0.03$ and $d_V=0.03$
contour by less than $10^{-3}$, and at least $98\%$ of the points of every contour by
less than one grid spacing in $\omega_i$. Larger displacements, of at most $0.09$, are confined to
tiny contour islands and to the narrow tips of pinched contours.

\begingroup
\setlength{\intextsep}{4pt plus 1pt minus 1pt}
\setlength{\floatsep}{4pt plus 1pt minus 1pt}
\setlength{\textfloatsep}{4pt plus 1pt minus 1pt}
\captionsetup{aboveskip=6pt,belowskip=0pt}

\begin{figure}
\centering
\includegraphics[width=\textwidth]{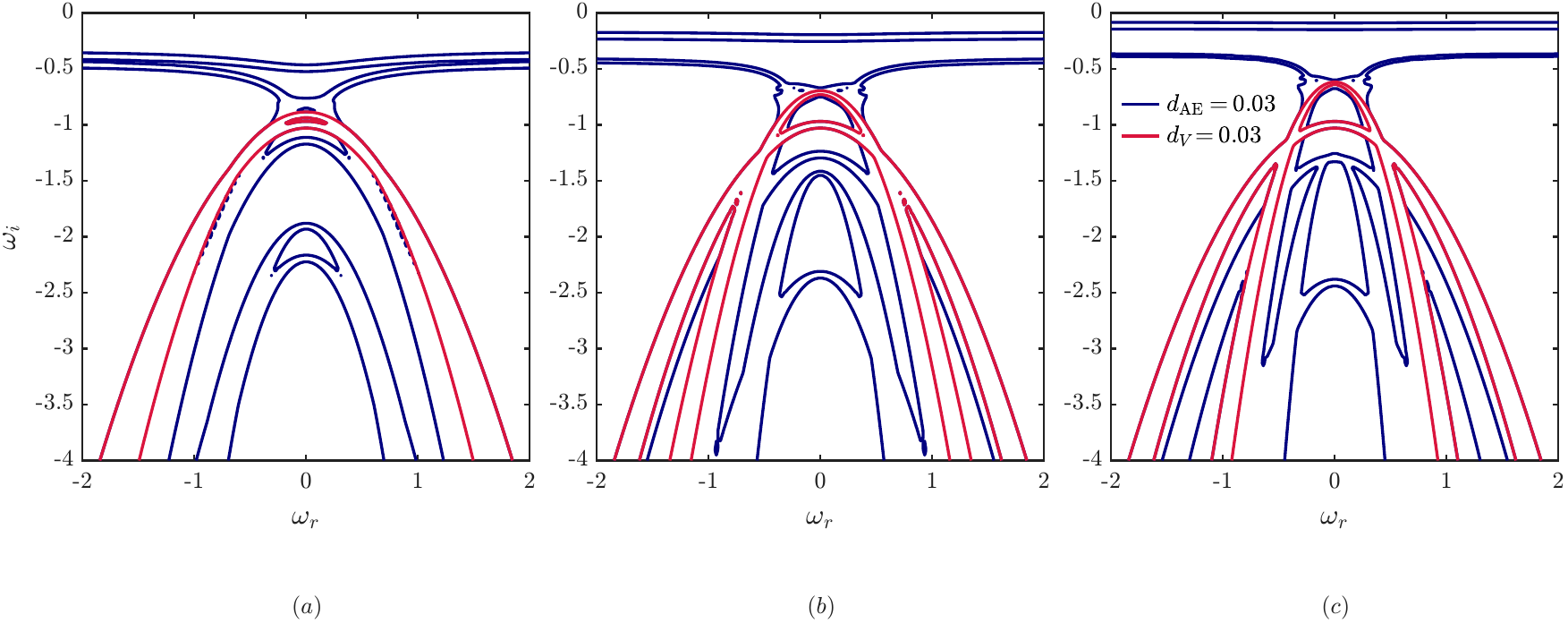}
\caption{$\Rey$-free contours $d_{\rm AE}=0.03$ (blue) and $d_V=0.03$ (red)
at $k_y=1$, on a $220\times180$ grid in $(\omega_r,\omega_i)$: (a) $M_1=1.1$, (b) $M_1=1.5$, (c) $M_1=2.0$.}
\label{fig:mach}
\vspace{6pt}
\centering
\includegraphics[width=\textwidth]{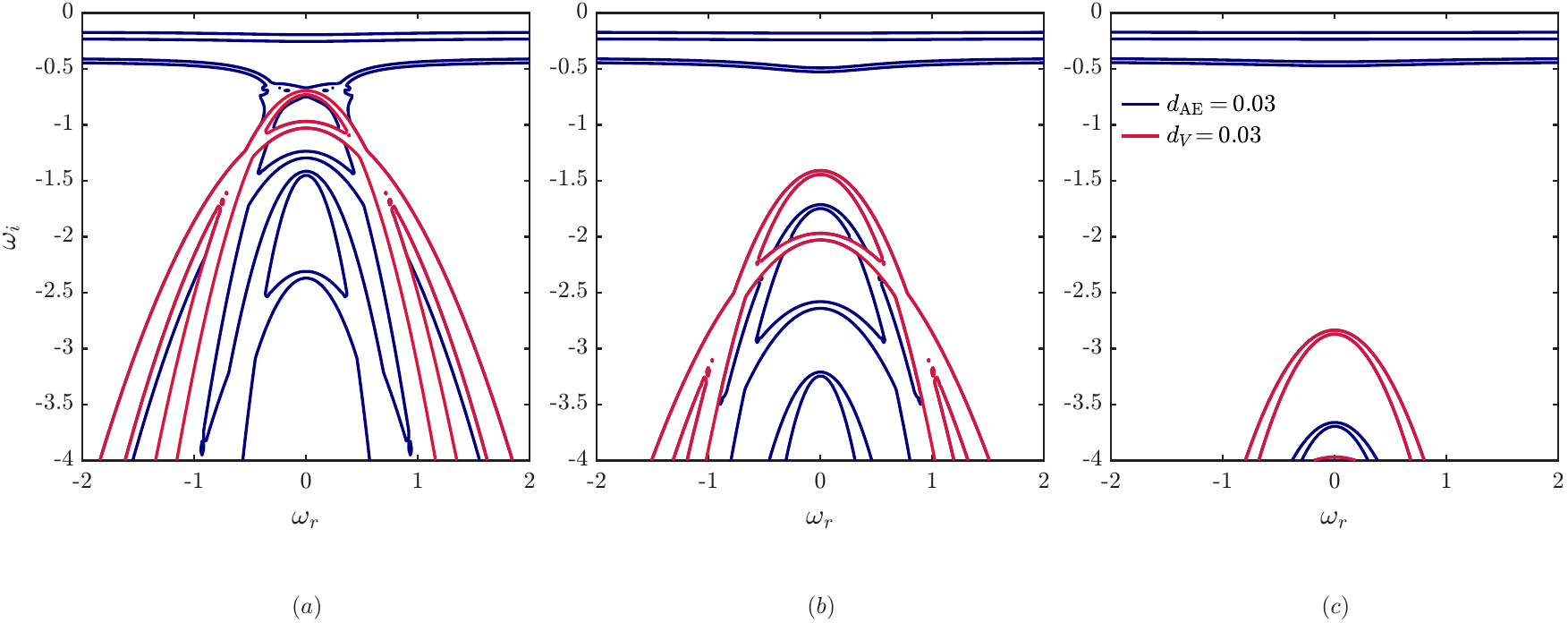}
\caption{$\Rey$-free discriminant contours $d_{\rm AE}=0.03$ (blue) and $d_V=0.03$ (red)
at $M_1=1.5$, on a $220\times180$ grid in $(\omega_r,\omega_i)$: (a) $k_z=0$ ($\kt=1$), (b) $k_z=1$ ($\kt=\sqrt2$), (c)
$k_z=\sqrt3$ ($\kt=2$), all at $k_y=1$.}
\label{fig:squire}
\end{figure}
\endgroup

A single discriminant spectral map solving the $\Rey$-free equations \eqref{eq:gs-vorticity}--\eqref{eq:gs-at}
at fixed $M_1,\gamma,\Pran,k_y$ determines the far-field branch structure
simultaneously at all Reynolds number values, so the map need be computed once per Mach
number and constitutive closure $(\gamma,\Pran,\mu(T))$, swept over $k_t$, rather
than once per $(\Rey,M_1)$ pair as the outer problem discussed by
\citet{Theofilis2026} would require; \S\ref{sec:theory-reynolds-collapse} makes this
precise and distinguishes it from the arbitrary choice of outer length underlying $\Rey$
itself.
Figure~\ref{fig:mach} shows this result for three weak-to-moderate shocks,
$M_1=1.1,1.5,2.0$, at fixed $k_y=1$. Their maximum-slope thickness
$\delta_s=(u_1-u_2)/\max_x|\D\ub|$ is $51.5$, $14.9$ and $11.2$ viscous lengths
$\ell_{\rm visc}^\ast$, against an upstream hard-sphere mean free path, consistent with
$\mu\propto T^{1/2}$, of $\ell_{\rm mfp}=\tfrac{16}{5}(\gamma/2\pi)^{1/2}M_1\,\ell_{\rm visc}^\ast$.
The layers are therefore about $28$, $6.0$ and $3.4$ mean free paths thick, respectively, and over the Mach
numbers $1.25$--$2$ of \S\ref{sec:pseudospectra} the thickness falls from $11.5$ to $3.4$ mean free paths.
Differences between continuum and kinetic shock structures grow with the Mach number, first and most visibly on the
upstream side of the layer \citep{LiepmannNarasimhaChahine1962}, so that at $M_1=2$ the continuum description is at the edge of its validity.
Results at $M_1=2$, here and in \S\ref{sec:pseudospectra}, are accordingly properties of the
Navier--Stokes--Fourier model rather than a quantitatively validated description of a real
monatomic shock. All three share a near-horizontal band of small
$d_{\rm AE}$ close to the real axis, essentially unchanged in position across the
Mach-number range studied. This feature is consistent with the near-neutral band already
visible in the spectra of \citet{DuckBalakumar1992}, though only qualitatively so, since
their spectra are computed in the original, primitive variables rather than the rotated,
decoupled acoustic--entropy/vortical form used here. Nested within a broader arch reaching toward more negative
$\omega_i$, the $d_V=0.03$ contour tracks the innermost $d_{\rm AE}$ loci closely at every
$M_1$. The arch is comparatively sparse at $M_1=1.1$ and becomes markedly richer, with more
nested sub-branches and more near-coincidences between the two branches, as $M_1$ increases
to $2.0$, i.e.\ as the downstream state departs further from the upstream one. Only
$\omega_i\le0$ is scanned here. The continuous spectra of both branches
lie strictly in $\omega_i<0$, by \eqref{eq:vorticity-dispersion} for the vortical branch and by \eqref{eq:chu-result} for the acoustic--entropy branch,
discrete vortical eigenvalues lie below the bound \eqref{eq:v-kt-bound}, and no unstable
acoustic--entropy eigenvalue appears in any of the full-layer computations of
\S\ref{sec:transient-results} and appendix~\ref{app:evp-validation}. The continuous
spectrum of the far field is thus strictly damped for every $\kt$, as figure~\ref{fig:mach} illustrates.

\subsection{Stabilization of the spectrum by the transverse wavenumber}
\label{sec:results-squire}

In \S\ref{sec:theory-theorem-consequence}--\S\ref{sec:theory-energy-ae} it was shown that
increasing the total transverse wavenumber $\kt=\sqrt{k_y^2+k_z^2}$ at fixed $k_y$, i.e.\ making a
disturbance genuinely three-dimensional, shifts the vortical continuous spectrum rigidly away
from the real axis by $(\mu_\infty/\rho_\infty)\kt^2$, while the acoustic--entropy continuous
spectrum remains strictly damped. Figure~\ref{fig:squire} illustrates both effects at $M_1=1.5$ and $k_y=1$ fixed, for $k_z=0,1,\sqrt3$, i.e. $\kt=1,\sqrt2,2$. 
By the covariance theorem the map for $(k_y,k_z)=(1,k_z)$ is the map at $k_y=\kt$ computed in the rotated frame.
It can be seen that the near-neutral locus of the acoustic--entropy branch is the least-damped feature
in every panel, sitting at $\omega_i\approx-0.17$ on the $d_{\rm AE}=0.03$
contour, offset from the exact plateau of \S\ref{sec:results-kt-sweep} and essentially $\kt$-invariant. By
contrast, the least-damped point of the vortical branch, the vertex of its downstream
continuous spectrum, retreats from $\omega_i=-0.713$ at $\kt=1$ to $-1.426$ at $\kt=\sqrt2$ and
$-2.853$ at $\kt=2$, exactly as $-(\mu_2/\rho_2)\kt^2$ in accordance with
\eqref{eq:vorticity-dispersion}.\footnote{The $d_V=0.03$ contours read $-0.70$, $-1.41$ and $-2.83$.} No oblique disturbance in
figure~\ref{fig:squire} is ever less damped than its $k_z=0$ counterpart at the same $k_y$.
This is precisely the pattern reported numerically, at fixed $\Rey=1000$ in fig.~4 of 
\citet{Theofilis2026} for the Reynolds-dependent outer problem. Figure~\ref{fig:squire} reproduces it
here for the continuous spectrum of the weak shock layers, exactly for the vortical branch and
as a strict damping bound for the acoustic--entropy branch, valid at all Reynolds numbers.

\subsection{Transverse-wavenumber sweep}
\label{sec:results-kt-sweep}

The dependence of the damping rate $\omega_i$ on $\kt$ for the vortical branch is explicitly
given by \eqref{eq:vorticity-dispersion}, while for
the acoustic--entropy branch the bound \eqref{eq:chu-result} does not by itself imply
monotonicity, as discussed in \S\ref{sec:theory-energy-ae}. Consequently, the least-damped point of the branch is tracked
numerically over a dense grid in $\kt$. With $\omega_j(\xi,\kt)$ the four roots of
$\mathbf A_0(\xi;\kt)\hat{\mathbf q}=\omega\mathbf C\hat{\mathbf q}$ (\S\ref{sec:theory-dispersion-ae}),
the relevant quantity is the supremum over $\xi$,
\begin{equation}
\omega_i^{\rm AE}(\kt)\equiv\sup_{\xi\in\mathbb R}\ \max_{j=1,\dots,4}\ {\rm Im}\,\omega_j(\xi,\kt),
\label{eq:kt-sweep-def}
\end{equation}
the spectral abscissa of the branch at that $\kt$, which need not be attained at $\xi=0$.
It is evaluated at $M_1=1.5$ for both far-field states on a uniform grid of 161 points,
that is 160 intervals of width $h\simeq0.016$, spanning $\kt\in(0,2.6]$. The sweep was
repeated on grids of 81 and 641 points with the same trend and plateau values. At each $\kt$ in the range where
the supremum is attained at finite $\xi$ it is located to $10^{-12}$.

On either far-field state, $\omega_i^{\rm AE}(\kt)$ is non-increasing on all three grids,
the largest positive step being of order $10^{-12}$, i.e.\ round-off.
Two regimes appear on both sides, separated by a finite crossover wavenumber $\kt^\ast$. Below $\kt^\ast$ the supremum is attained
at $\xi=0$, on the decoupled streamwise root of \eqref{eq:ae-dispersion-matrix}, and
$\omega_i^{\rm AE}(\kt)=-(\mu_\infty/\rho_\infty)\kt^2$, as for the vortical branch. Beyond
$\kt^\ast$ it is attained as $\xi\to\pm\infty$, an accumulation point of the continuous spectrum,
on a coupled root of the $(\rh,\vpar,\Th)$
block, and $\omega_i^{\rm AE}(\kt)$ stays on an exact, $\kt$-independent plateau out to
$\kt=2.6$, which the numerical values, computed for $|\xi|\le200$, reproduce to about $10^{-6}$; unlike the vortical damping, which grows as $\kt^2$, the acoustic--entropy damping
saturates. The crossover occurs where the plateau meets the vortical damping, so that
$\kt^{\ast2}=|\omega_i^{\rm AE}(\kt\to\infty)|\,\rho_\infty/\mu_\infty$.
Table~\ref{tab:kt-sweep} gives the crossover wavenumber $\kt^\ast$ at $M_1=1.5$ and the plateau value
at five Mach numbers spanning those used in \S\ref{sec:results} and \S\ref{sec:pseudospectra}, at
all of which the same non-increasing trend and finite-plateau structure hold.
This computation also calibrates the offset contours of
figures~\ref{fig:mach}--\ref{fig:squire}: the near-neutral acoustic--entropy locus read off the
$d_{\rm AE}=0.03$ contours at $M_1=1.5$, $\omega_i\approx-0.17$, lies above the exact plateau,
$-0.200$, by the finite offset of the contour from the true zero set.

\begin{table}
\centering
\caption{Acoustic--entropy branch: crossover wavenumber $\kt^\ast$ at $M_1=1.5$ and plateau value
$\omega_i^{\rm AE}(\kt\to\infty)$ at each Mach number, by far-field state.}
\label{tab:kt-sweep}
\begin{tabular}{lcccccc}
\toprule
\shortstack{far-field\\state} & $\kt^\ast$ & \multicolumn{5}{c}{$M_1$}\\
\cmidrule(l){3-7}
 & & $1.1$ & $1.25$ & $1.5$ & $1.75$ & $2.0$\\
\midrule
upstream (cold) & $0.447$ & $-0.3719$ & $-0.2880$ & $-0.2000$ & $-0.1469$ & $-0.1125$\\
downstream (hot) & $0.767$ & $-0.4481$ & $-0.4399$ & $-0.4192$ & $-0.3951$ & $-0.3707$\\
\bottomrule
\end{tabular}
\end{table}

The plateau has a closed form. As $|\xi|\to\infty$ the viscous terms of
\eqref{eq:ae-dispersion-matrix} dominate and one root of the quartic \eqref{eq:ae-dispersion-det},
associated with the longitudinal viscous mode, tends to
$\omega\simeq u_\infty\xi-\mathrm i\,p_\infty/(\lambda_\infty+\tfrac43\mu_\infty)$, where
$p_\infty=\rho_\infty T_\infty/(\gamma M_1^2)$ is the far-field pressure. For $\lambda_\infty=0$
the plateau is therefore
\begin{equation}
\omega_i^{\rm AE}(\kt\to\infty)=-\frac{3\,\rho_\infty T_\infty}{4\gamma M_1^2\,\mu_\infty}
=-\frac{3\,\mu_\infty\rho_\infty}{4\gamma M_1^2}\quad(\mu\propto T^{1/2}),
\label{eq:ae-plateau}
\end{equation}
independent of $\kt$, which is why the plateau is flat at $-0.45/M_1^2$ at the upstream state, and \eqref{eq:ae-plateau} reproduces every entry of table~\ref{tab:kt-sweep}.

\section{Non-modal results}
\label{sec:pseudospectra}

Linear non-modal analysis has been performed by numerical solution of three linear initial-value problems governed by equation~\eqref{eqn:LST_IVP}, for the uniform upstream gas, the uniform downstream gas and the shock layer in between.
In each uniform gas the acoustic--entropy disturbance satisfies a $4\times4$ constant-coefficient system at each wavenumber pair $(\xi,k_y)$.
Its propagator is evaluated numerically as a function of time, and optimal transient growth in the Euclidean norm is computed from the largest singular value of the propagator \citep{Farrell1988,ButlerFarrell1992,ReddyHenningson1993,Trefethen1993}.
The resolvent of the same generator is evaluated on a grid of trial eigenvalues in the complex plane, and the contours of its norm define the pseudospectra.
In the shock layer, the full, $x$-dependent acoustic--entropy part of the operator $\mathbf A$ of equation~\eqref{eqn:LST_IVP} over the Gilbarg--Paolucci base flow is discretized by Chebyshev collocation on a truncated
domain, and the norm of its propagator is computed numerically over a tapered window, the discretization being verified in appendix~\ref{app:evp-validation}.
Comparing the results of the three initial value problems identifies transient growth contributions of the uniform gases and the shock layer itself.

\subsection{The far-field generator, disturbance norm, and resolvent}
\label{sec:transient-setup}

The exact far-field dispersion relation of the acoustic--entropy branch
\eqref{eq:ae-dispersion-det} was obtained in \S\ref{sec:theory-dispersion-ae} from the
generalized eigenvalue problem $\mathbf A_0(\xi;k_y)\hat{\mathbf q}=\omega\mathbf C
\hat{\mathbf q}$, with
$\hat{\mathbf q}=(\rh,\uh,\vpar,\Th)^{\rm T}$ and  $\mathbf
D=\mathrm{diag}(1,\mathrm i\rho_\infty,\mathrm i\rho_\infty,\mathrm i\rho_\infty)$, and
$\mathbf A_0$ was presented explicitly in \S\ref{sec:theory-dispersion-ae},
eq.~\eqref{eq:ae-dispersion-matrix-explicit} with $\Delta\to u_\infty\xi$. In the limit of a uniform gas, spatial differentiation
simplifies to $\partial/\partial x\to\mathrm i\xi$, and equation~\eqref{eqn:LST_IVP} becomes
the linear \emph{initial-value} problem obtained by substituting the
plane-wave ansatz into \eqref{eq:gs-at},
\begin{equation}
\frac{\partial\hat{\mathbf q}}{\partial t}=\mathbf A_{\rm gen}(\xi,k_y)\,\hat{\mathbf q},\qquad
\mathbf A_{\rm gen}(\xi,k_y)\equiv-\mathrm i\,\mathbf C^{-1}\mathbf A_0(\xi,k_y).
\label{eq:ae-generator}
\end{equation}
The eigenvalues are exactly $-\mathrm i\,\omega_j(\xi,k_y)$, $j=1,\dots,4$, the same four
roots of \eqref{eq:ae-dispersion-det} already computed and verified against the
companion-matrix diagnostics of \S\ref{sec:theory-companion-discriminant} in
figure~\ref{fig:dispersion-check}(b) of appendix~\ref{app:dispersion-check}. Solution of the initial-value problem \eqref{eq:ae-generator} examines whether, although every eigenvalue is stable,
non-normality allows $\|\hat{\mathbf q}(t)\|$ to grow transiently before it decays asymptotically in time, a
mechanism inaccessible to modal analysis \citep{Farrell1988,ButlerFarrell1992,ReddyHenningson1993,Trefethen1993}.

Disturbance amplitude is measured here in the plain Euclidean norm on the primitive
state, $\|\hat{\mathbf q}\|^2=|\rh|^2+|\uh|^2+|\vpar|^2+|\Th|^2$, the norm underlying
the pseudospectra of a matrix \citep{TrefethenEmbree2005}. The optimal transient growth at time $t$
is then
\begin{equation}
G(t)\equiv\sup_{\hat{\mathbf q}_0\neq\mathbf 0}\frac{\|\hat{\mathbf q}(t)\|^2}{\|\hat{\mathbf q}_0\|^2}
=\big\|\exp(t\,\mathbf A_{\rm gen})\big\|_2^2,
\label{eq:transient-growth-def}
\end{equation}
the squared operator 2-norm of the propagator, which is attained by the
corresponding right singular vector of $\exp(t\mathbf A_{\rm gen})$ as initial condition.
The evaluation of this norm by singular value decomposition, the right singular vector giving the optimal initial condition, follows 
\citet{ButlerFarrell1992} and \citet{ReddyHenningson1993},
who applied it in a truncated eigenfunction basis. By contrast, here no modal basis is required, since the decomposition is applied to the propagator itself.
The same optimum can be obtained without forming the propagator, by iterating the direct problem forward in time and its adjoint backward,
the methodology established for optimal perturbations and receptivity by \citet{LuchiniBottaro1998}; see also \citet{LuchiniBottaro2014}.
This direct--adjoint looping was used as an additional, independent confirmation of the propagator results that follow, with implicit time integration, Richardson extrapolation in the time step
and the Euclidean norm of the discretization. In the cold and hot uniform far fields it reproduces \eqref{eq:transient-growth-def} to $3\times10^{-14}$ at all four Mach numbers and for
$t=0.5$--$4$. For the full shock-layer generator at $M_1=2$ direct--adjoint looping reproduces the SVD-obtained results to within $2\times10^{-4}$ at $t=1$, and $7\times10^{-6}$ at the optimal time.
Boundedness of $G$ at the origin $(\xi,k_y)=(0,0)$, where the generator degenerates to zero, is discussed in
appendix~\ref{app:origin-boundedness}.

The Euclidean norm used here differs from the disturbance energy of \citet{Chu1965}
\eqref{eq:chu-energy} of \S\ref{sec:theory-energy-ae}. In the Chu norm, whose weights \eqref{eq:chu-weights}
make the pressure-work exchange between every pair of fields cancel exactly, the same far-field
generator, written as $\mathbf A_{\rm Chu}=\mathbf W^{1/2}\mathbf A_{\rm gen}\mathbf W^{-1/2}$ with $\mathbf W={\rm diag}(\beta_\rho,\rho_\infty,\rho_\infty,\rho_\infty\beta_T)$,
has been found to be \emph{purely dissipative}, $\max{\rm spec}\big((\mathbf A_{\rm Chu}+\mathbf A_{\rm Chu}^{\rm H})/2\big)\le0$ with ${\rm H}$ the conjugate transpose, to machine precision over a fine $(\xi,k_y)$-grid at every Mach
number and on both sides, leading to $G_{\rm Chu}(t)\equiv1$. It should be noted in this context that the results of the two norms are not mutually conflicting. The Chu norm is the one in which the acoustic--entropy disturbance energy is non-increasing,
the property used for the bound of \S\ref{sec:theory-energy-ae}, while the Euclidean norm
exposes the non-normality intrinsic to the primitive variables. The transient growth values reported below are
therefore those of the Euclidean norm.

\subsection{Pseudospectra of the hot-side generator}
\label{sec:pseudospectra-results}

As an analysis methodology complementary to the transient growth \eqref{eq:transient-growth-def}, the $\varepsilon$-pseudospectrum \citep{Trefethen1993}, defined by
\begin{equation}
\sigma_\varepsilon(\mathbf A_{\rm gen})\equiv\Big\{s\in\mathbb C:\ \big\|(s\mathbf
I-\mathbf A_{\rm gen})^{-1}\big\|_2\ge\varepsilon^{-1}\Big\},
\label{eq:pseudospectrum-def}
\end{equation}
has been computed. Its computation relies on the resolvent norm $\|(s\mathbf I-\mathbf A_{\rm gen})^{-1}\|_2$, and it reduces to the ordinary spectrum as $\varepsilon\to0$. For a non-normal operator the pseudospectrum
inflates into open regions at finite $\varepsilon$, and its area and reach into the upper
half-plane are controlled by exactly the same non-normality responsible for
\eqref{eq:transient-growth-def} \citep{TrefethenEmbree2005}. Since the eigenvalues of $\mathbf A_{\rm gen}$
are $s_j=-\mathrm i\,\omega_j(\xi,k_y)$, i.e.\ $\omega_j=\mathrm
is_j$, the pseudospectra below are plotted directly in the same $(\omega_r,\omega_i)$
eigenvalue plane used throughout \S\ref{sec:analytical} and \S\ref{sec:results}, by evaluating
the resolvent of \eqref{eq:pseudospectrum-def} at $s=-\mathrm i\omega$ on a grid of
$\omega=\omega_r+\mathrm i\omega_i$.

\begin{figure}[!b]
\centering
\includegraphics[width=\textwidth]{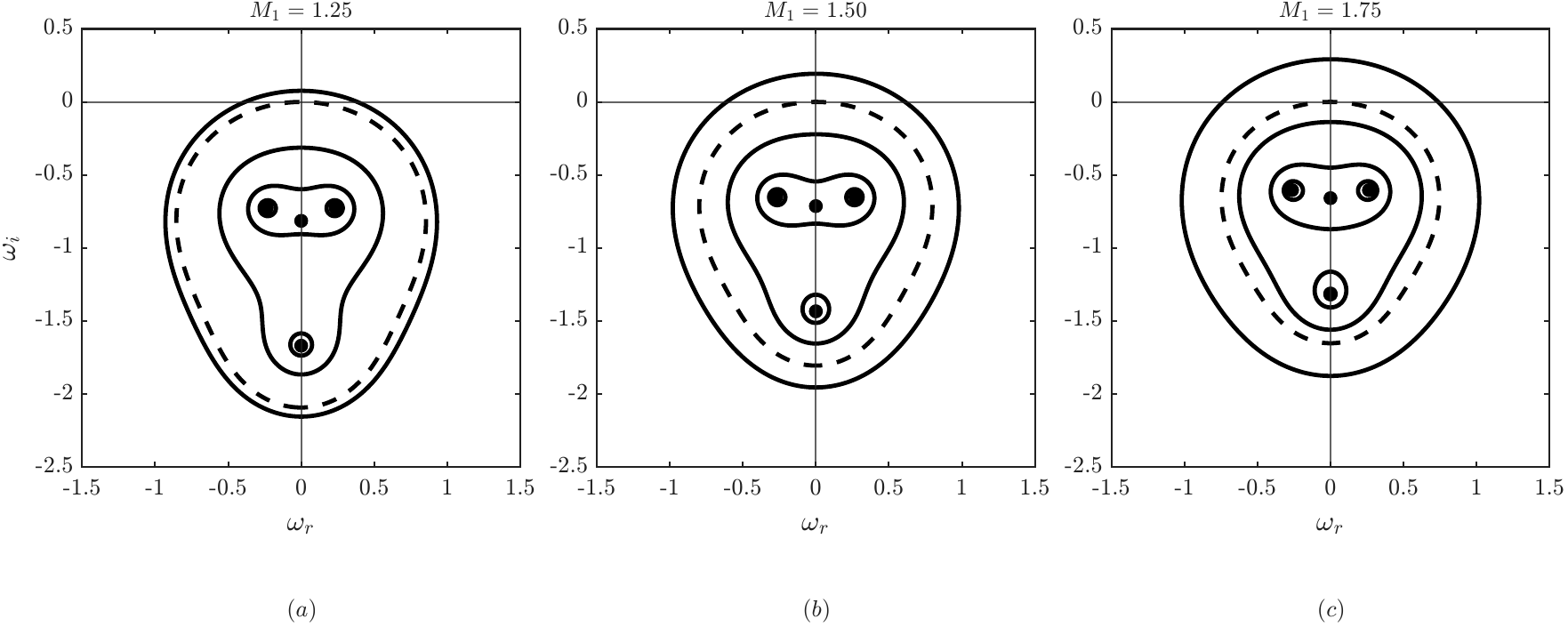}
\caption{$\varepsilon$-pseudospectra of $\mathbf A_{\rm gen}$ at $(\xi,k_y)=(0,1)$, downstream
(hot) far-field state, for (a) $M_1=1.25$, (b) $M_1=1.5$, (c) $M_1=1.75$. Filled circles, eigenvalues of $\mathbf A_{\rm gen}$. Dashed line, $\varepsilon=\varepsilon_c$.}
\label{fig:pseudospectra}
\end{figure}

Figure~\ref{fig:pseudospectra} shows the $\varepsilon$-pseudospectra \eqref{eq:pseudospectrum-def} of
$\mathbf A_{\rm gen}$ at $(\xi,k_y)=(0,1)$ in the downstream (hot) far-field state, for
$M_1=1.25$, $1.5$ and $1.75$.
Only the hot side is shown because it is
where the effects of non-normality are strongest, e.g.\ $G_{\max}\approx1.84$ downstream against $1.11$ upstream at $M_1=1.5$. Each panel contains the
nested (outermost to innermost, the same four levels in every panel) boundaries of $\sigma_\varepsilon$ 
for $\varepsilon=0.3,0.1,0.03,0.01$. Plotted are the dashed
boundary $\varepsilon=\varepsilon_c$ defined below and the
four eigenvalues of $\mathbf A_{\rm gen}$ as filled circles, namely a complex-conjugate pair, the stationary
eigenvalue between them and a more strongly damped stationary eigenvalue. The four eigenvalues are the four roots of the exact far-field dispersion
relation of \S\ref{sec:analytical-dispersion} at this wavenumber pair, that is, points of the acoustic--entropy continuous spectrum, and not discrete eigenvalues of the shock layer.

It should be noted that here $\varepsilon$ denotes the norm of a perturbation of the generator, $\|\mathbf F\|_2\le\varepsilon$ in $\mathbf A_{\rm gen}+\mathbf F$, and not the amplitude of the
disturbance, so that the linearization is exact for every $\varepsilon$.
All four eigenvalues lie below the line $\omega_i=0$ at every Mach number, so the generator is modally
stable. The pseudospectra nevertheless tell a different story. A normal operator would have
pseudospectra made of disks of radius $\varepsilon$ around each eigenvalue, whereas here the contours
of the larger values of $\varepsilon$ enclose all four eigenvalues in a single region far larger than such
disks. The smallest perturbation that places an eigenvalue in the upper half-plane is the complex stability radius
$\varepsilon_c=\min_{\omega\in\mathbb R}\sigma_{\min}(\mathrm i\omega\mathbf I-\mathbf A_{\rm gen})$, whose boundary is the dashed contour touching $\omega_i=0$.
It is $\varepsilon_c=0.25$, $0.19$, $0.15$ at $M_1=1.25$, $1.5$, $1.75$, that is $11\%$, $8\%$ and $6\%$ of $\|\mathbf A_{\rm gen}\|_2=2.31$, $2.40$, $2.64$,
and it decreases with $M_1$. Perturbations of $6$--$11\%$ of $\|\mathbf A_{\rm gen}\|_2$ therefore suffice to move an eigenvalue of the hot-side generator into the unstable half-plane, and the outermost contour shown, $\varepsilon=0.3>\varepsilon_c$,
reaches into the upper half-plane at every Mach number. This is the pseudospectral expression of the same
non-normality that produces transient growth $G(t)>1$ in a modally stable system. It is fair to characterize such transient growth
as moderate, as also indicated by the Kreiss constant,
$\mathcal K=\sup_{\mathrm{Re}\,s>0}\mathrm{Re}\,s\,\|(s\mathbf I-\mathbf A_{\rm gen})^{-1}\|_2=1.03$, $1.13$, $1.26$, a lower bound for $\sup_t\|e^{t\mathbf A_{\rm gen}}\|_2=\sqrt{G_{\max}}=1.13$, $1.36$, $1.61$. The same transient growth is
quantified next, in the time domain.

\subsection{Transient growth in the far fields and in the shock layer}
\label{sec:transient-results}

\begin{figure}[!b]
\centering
\begin{minipage}[c]{0.4\textwidth}
\includegraphics[width=\textwidth]{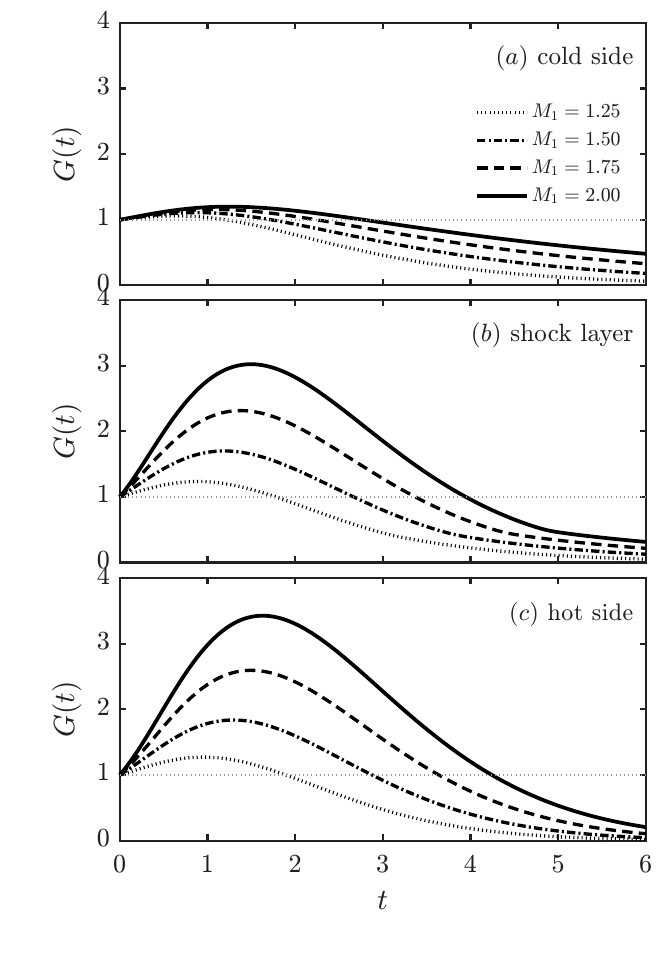}
\end{minipage}%
\begin{minipage}[c]{0.6\textwidth}
\includegraphics[width=\textwidth]{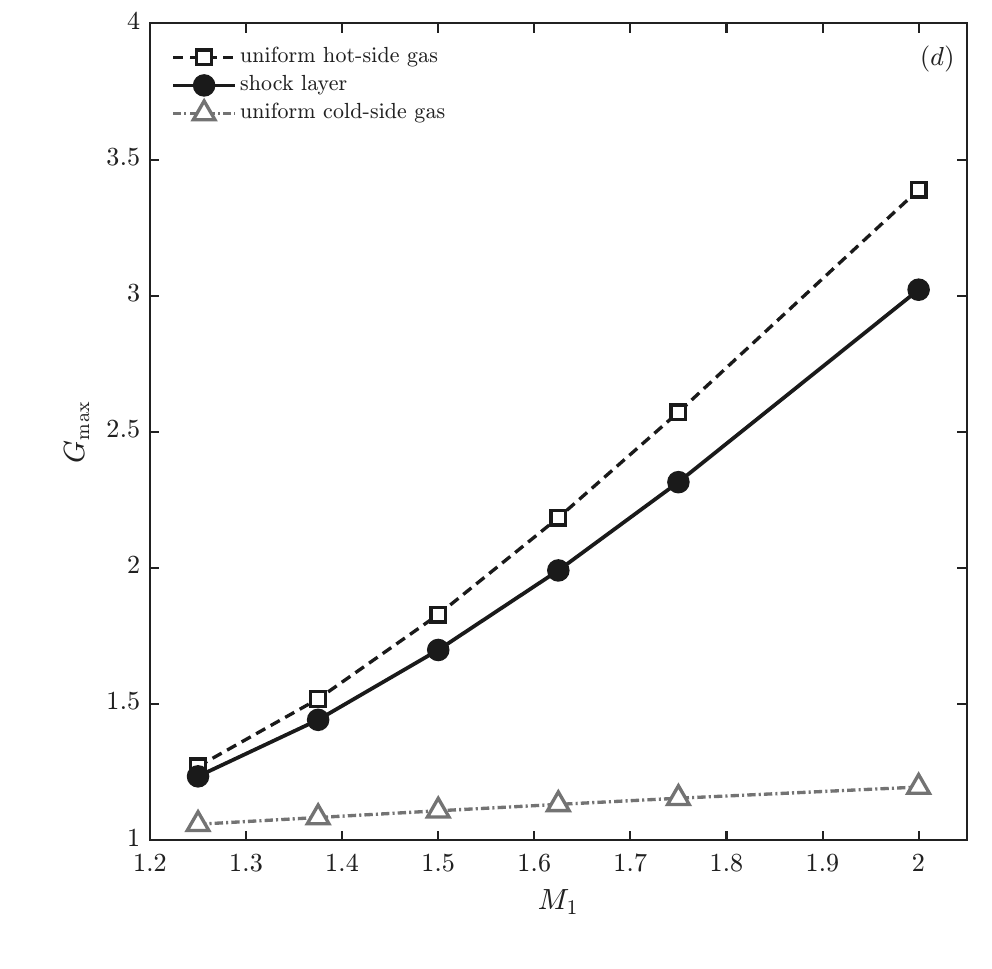}
\end{minipage}
\caption{(a)--(c) $G(t)$ at $(\xi,k_y)=(0,1)$ for $M_1=1.25,1.5,1.75,2$ (dotted, dash-dot,
dashed, solid): (a) cold side, (b) shock layer, window $(\tfrac23\delta_s,\tfrac53\delta_s)$, (c) hot
side. (d) Shock-local $G_{\max}$ against $M_1$: shock layer (filled circles), uniform hot-side
(open squares) and cold-side (open triangles) gas.}
\label{fig:transient-growth}
\end{figure}

Figure~\ref{fig:transient-growth}(a)--(c) reports $G(t)$ computed from \eqref{eq:transient-growth-def} at
the wavenumber pair $(\xi,k_y)=(0,1)$, with $\xi$ the streamwise wavenumber of \S\ref{sec:analytical-dispersion}, for
$M_1=1.25,1.5,1.75,2.0$, in three settings: (a) the uniform upstream
(cold) gas, (b) the shock layer itself, and (c) the uniform downstream (hot) gas, taking the two uniform gases first. In both far
fields the spectrum of $\mathbf A_{\rm gen}$, expressed through $\omega=\mathrm is$, lies strictly in the lower half-plane $\omega_i<0$, yet $G(t)>1$ over an intermediate range of $t$ at every Mach number.

The spectra differ in character on the two sides. Upstream all four eigenvalues are purely imaginary at every
Mach number tested, i.e.\ stationary and decaying without oscillation, for example
$\omega\in\{-0.225\mathrm i,\,-1.000\mathrm i,\,-1.267\mathrm i,\,-2.342\mathrm i\}$ at $M_1=1.5$. Downstream
a pair of oscillatory, complex-conjugate eigenvalues, $\omega=\pm0.270-0.650\mathrm i$ at $M_1=1.5$, accompanies two stationary ones,
$\omega\in\{-1.434\mathrm i,\,-0.713\mathrm i\}$. At $\xi=0$ the eigenvalue $-\mathrm i\mu_\infty k_y^2/\rho_\infty$ of the streamwise velocity decouples, and
the other three comprise a thermal eigenvalue, which is always stationary, and an acoustic pair, which is oscillatory only if
$\nu k_y/c$ lies below a critical value close to unity, with $\nu=\mu_\infty/\rho_\infty$ and $c=T_\infty^{1/2}/M_1$ the sound speed. Transverse sound waves therefore
propagate only if $k_y\lesssim c/\nu$, i.e.\ if their wavelength exceeds several times the viscous length $\nu/c$, which is of the order of the mean free path.
At $k_y=1$ the ratio $\nu k_y/c$ equals $M_1$ upstream, where sound is overdamped, and is about $0.9$ downstream, where the
denser gas supports weakly damped sound waves.
A damped
spectrum thus supports transient growth in the Euclidean norm at this finite, $O(1)$ wavenumber, a
mechanism whose boundedness at the origin is shown in appendix~\ref{app:origin-boundedness}.
The values of $G_{\max}$ quoted below refer to this wavenumber pair and are not global maxima, since larger values are approached as the wavenumbers tend to zero (appendix~\ref{app:origin-boundedness}).

Upstream, figure~\ref{fig:transient-growth}(a), the growth is weak:
$G_{\max}=1.06$, $1.11$, $1.16$, $1.20$ at $M_1=1.25,1.5,1.75,2.0$ ($t_{\rm opt}=0.65$,
$0.87$, $1.07$, $1.25$), barely above unity and rising roughly linearly with $M_1$. The
optimal initial disturbance is already mostly density ($78\%$ of the norm at $M_1=1.5$, the
rest $\vpar$) and relaxes by $t_{\rm opt}$ into an almost pure density perturbation,
leaving little kinetic-to-acoustic conversion to drive growth.

Downstream, figure~\ref{fig:transient-growth}(c), the growth is substantially stronger and
increases markedly with Mach number, $G_{\max}=1.27$, $1.84$, $2.60$, $3.43$ ($t_{\rm
opt}=0.90$, $1.36$, $1.51$, $1.63$), reproduced to within $2\%$ by the
quadratic
\begin{equation}
G_{\max}(M_1)\ \approx\ 1.067\,M_1^2-0.579\,M_1+0.323.
\label{eq:gmax-mach-fit}
\end{equation}
With four sampled Mach numbers, this form is only an empirical interpolant over
$M_1\in[1.25,2.0]$. At $t=0$ the optimum is dominated by $\vpar$ ($66\%$ of the norm at
$M_1=1.5$, $84\%$ at $M_1=2.0$), which relaxes by $t_{\rm opt}$ into an almost pure density
perturbation ($\gtrsim0.99$ of the norm at every Mach number). In other words, the growth is a transient
conversion of cross-stream velocity into density/pressure fluctuation. In both far fields $\uh$
and $\Th$ remain negligible and $\xi=0$ is the optimal streamwise wavenumber at this $k_y$, consistent with
appendix~\ref{app:origin-boundedness}. The pseudospectra of \S\ref{sec:pseudospectra-results} shown in figure~\ref{fig:pseudospectra} show the corresponding distortion of the resolvent norm, which also grows with Mach number.

The shock layer, figure~\ref{fig:transient-growth}(b), is computed with the full, $x$-dependent V and AE operators,
discretized as in appendix~\ref{app:evp-validation}, at the four Mach numbers above and at $M_1=1.375$ and $1.625$, six in all.
Their least-damped spectrum consists of boundary-localized
eigenvalues that discretize the continuous spectrum, so no small modal basis exists and the propagator approach of
\S\ref{sec:transient-setup} is used. The V branch shows no transient growth, $G(t)\le1$ for all $t$. The AE generator is modally stable at all six Mach numbers,
its spectral abscissa rising from $-0.32$ at $M_1=1.25$ to $-0.13$ at $M_1=2$, the numerical evidence for the absence of
unstable acoustic--entropy eigenvalues cited in \S\ref{sec:theory-energy-ae}, yet it supports transient growth. Evaluated over the whole truncated domain, this growth
peaks at the open outflow boundary and moves with it, a truncation artifact, so initial
disturbances and response are restricted to a window with a smooth cosine taper (appendix~\ref{app:evp-validation}). A window wide enough to clear the shock merely
reproduces the downstream far-field optimum of figure~\ref{fig:transient-growth}(c), because it contains mostly post-shock gas
and does not isolate the shock (appendix~\ref{app:evp-validation}). The window is therefore tied to the shock thickness, $\delta_s=(u_1-u_2)/\max_x|\D\ub|$,
$(c_1,c_2)=(\tfrac23\delta_s,\tfrac53\delta_s)$, and the same window and discretization are applied to two comparison operators with the base flow frozen at its upstream
and at its downstream state. The resulting shock-layer $G_{\max}$, figures~\ref{fig:transient-growth}(b) and (d), is
$1.234$, $1.442$, $1.700$, $1.991$, $2.315$, $3.022$ at the six Mach numbers, between the values of the two comparison operators
and, for this window, consistently $82$--$83\%$ of the way from upstream to downstream. The viscous shock
structure thus supplies no transient amplification beyond that of the post-shock gas, and the growth
it supports is drawn from its downstream side.

\begin{figure}
\centering
\includegraphics[width=\textwidth]{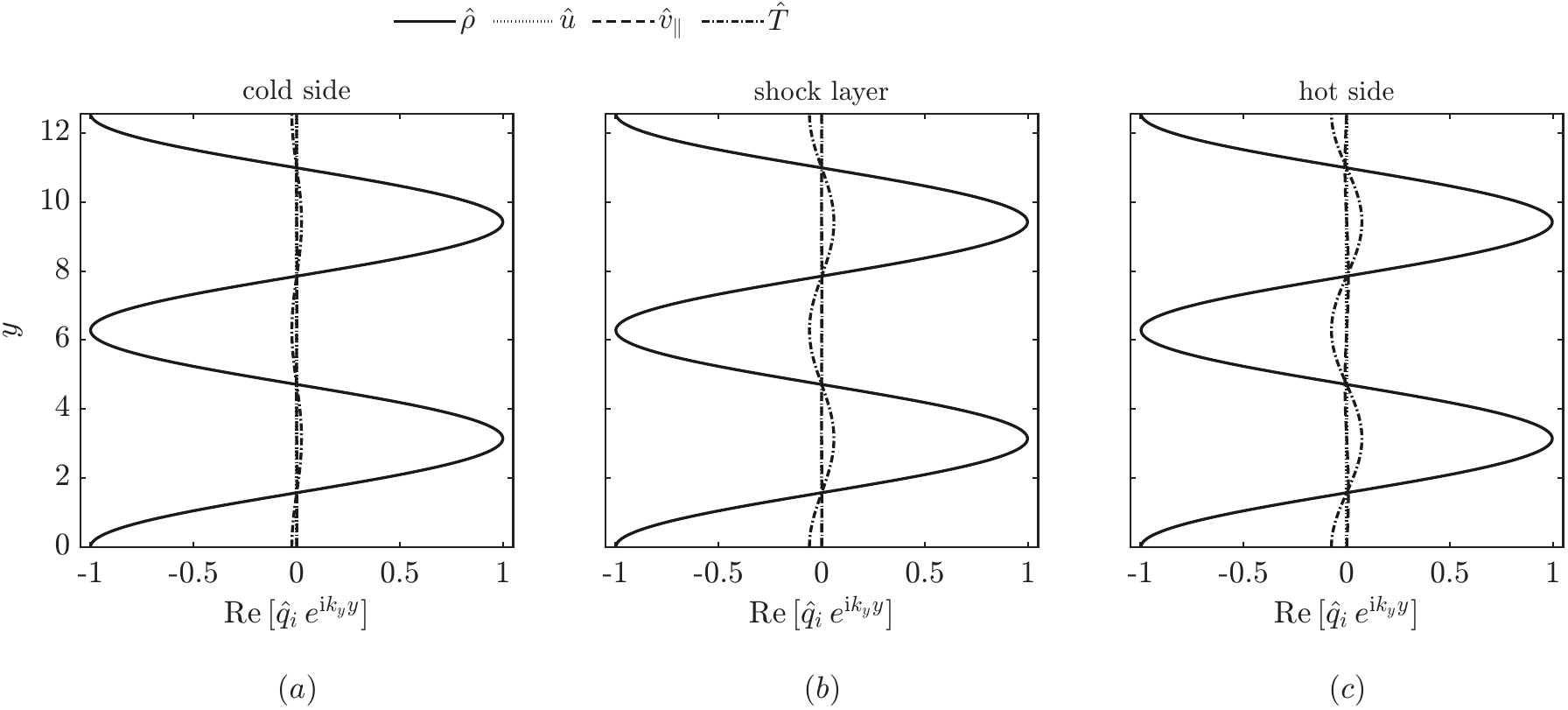}
\caption{${\rm Re}[\hat q_i(t_{\rm opt})\,e^{\mathrm ik_yy}]$ for
$\rh,\uh,\vpar,\Th$ at $(M_1,\xi,k_y)=(1.5,0,1)$. (a) Upstream far field.
(b) Shock layer, at the density peak $x=9.2$ of its shock-local optimum. (c) Downstream
far field.}
\label{fig:optimal-perturbation}
\end{figure}

The structure of the optimal disturbances is shown in figure~\ref{fig:optimal-perturbation} at $M_1=1.5$. In both far fields the optimum has $\xi=0$, so it is uniform in $x$ and varies only as $e^{\mathrm ik_yy}$, and figure~\ref{fig:optimal-perturbation}(a,c) therefore plots
$\mathrm{Re}\,[\hat q_i(t_{\rm opt})\,e^{\mathrm ik_yy}]$ against $y$ (the imaginary part has
the same shape, shifted by a quarter wavelength). The density perturbation $\rh$ and, much more
weakly, the temperature perturbation $\Th$ trace clean sinusoids, $\uh$ vanishes identically ($\xi=0$ decouples it)
and $\vpar$ is negligible by $t_{\rm opt}$. The only visible difference is the temperature
part, $\Th/\rh\approx0.023$ upstream against $0.074$ downstream, the same hot-side
amplification seen in $G_{\max}$.

\begin{figure}[!t]
\centering
\includegraphics[width=0.42\textwidth]{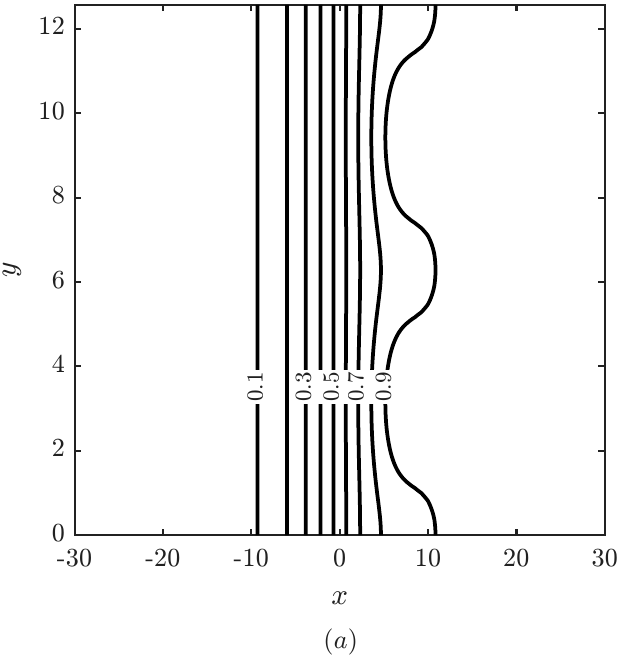}\hfill
\includegraphics[width=0.42\textwidth]{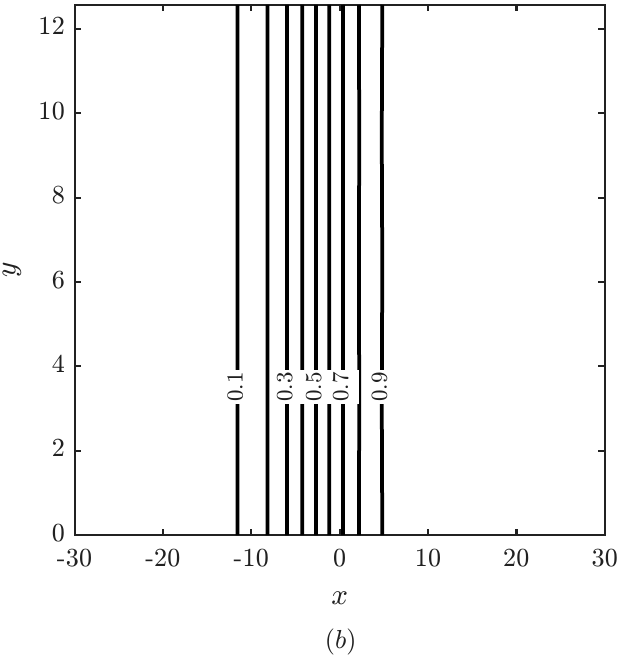}
\caption{Base flow plus shock-local optimal disturbance at $t=t_{\rm opt}$,
$(M_1,\gamma,\Pran,k_y)=(1.5,\tfrac53,\tfrac23,1)$, $t=0$ density peak $5\%$ of the jump;
contours at $0.1,\ldots,0.9$ of each jump. (a) $(\rho-\rho_1)/(\rho_2-\rho_1)$. (b)
$(u_1-u)/(u_1-u_2)$.}
\label{fig:tg-totalfield}
\end{figure}

The shock-layer optimum has the same structure at every Mach number. At $t=0$ it combines
density and $\vpar$ ($38\%$ and $58\%$ of the norm at $M_1=1.5$), and by $t_{\rm opt}$ ($1.19$ at $M_1=1.5$) it has
relaxed into an almost pure density perturbation, over $99\%$ of the norm, as in both far fields.
Its phase is uniform across the layer. At its density peak ($x=9.2$ at $M_1=1.5$) it is locally the far-field
structure, figure~\ref{fig:optimal-perturbation}(b), with $\Th/\rh=0.060$ between the
upstream and downstream values and closer to the latter, as is its growth.

Figure~\ref{fig:tg-totalfield} shows the effect of linear superposition of perturbations to the steady base flow
of the shock layer, $q(x,y)=\bar q(x)+a\,{\rm Re}[\hat q(x,t_{\rm opt})\,e^{\mathrm ik_yy}]$, 
with $a$ such that the $t=0$ density peak is $5\%$ of the
density jump. The $0.8$ and $0.9$ density contours ripple with the transverse wavelength
$2\pi/k_y$, by about $\pm3$ viscous lengths for the $0.9$ contour, while the contours from $0.1$
to $0.7$, spanning the steep part of the profile, remain straight, as do all velocity contours
(the velocity perturbation stays below $0.2\%$ of its jump, against some $11\%$ for density). The disturbance thus ripples the density on the post-shock side of the layer
without displacing the shock front, whether that front is marked by velocity or by the steep
part of the density profile. The picture holds only for small amplitudes: with a $t=0$ density peak of $8\%$ of the jump,
the $0.9$ contour already pinches off into closed islands.

Physically, then, a transversely periodic disturbance already present in the free stream is only
weakly amplified before it reaches the shock, at most $20\%$ in the squared Euclidean norm for $M_1\le2$. One of the same wavenumber
generated behind it, by shock corrugation or shock--turbulence interaction, can instead be amplified up to about $3.4$-fold, at $M_1=2$, before it too relaxes into a density
perturbation, and the shock structure in between adds nothing to this. None of this models
transmission through the shock, which has been treated for an idealized discontinuity by linear
interaction analyses \citep{Ribner1954,McKenzieWestphal1968} and, for the stagnation streamline
behind a blunt-body shock, by the analysis of \citet{MilmanKarp2026}. Interestingly, that analysis also finds the post-shock flow to be the more amplifying, and coupling that inviscid result to the viscous non-normal evolutions documented here
constitutes a natural avenue for future work.

\section{Conclusions}
\label{sec:conclusions}

The central result of this work is the rotational covariance theorem and its Squire-like consequences. For any base
flow depending on the shock-normal coordinate alone, the reduction is exact and Reynolds-free, and it leaves a two-dimensional
acoustic--entropy problem and a scalar vortical equation with a closed-form dispersion relation. One computation therefore
serves every orientation of the transverse wavenumber vector and every Reynolds number.

Using the viscous length scale of the shock, the steady laminar base flow and the three-dimensional
small-amplitude perturbations of a shock layer have been formulated in $\Rey$-free inner variables.
In these variables every three-dimensional eigenmode of a base flow depending on the shock-normal
coordinate alone, with zero transverse velocity, decouples exactly into a two-dimensional in-plane
acoustic--entropy (AE) branch carrying the entire dilatation and a one-dimensional out-of-plane
vortical (V) branch governed by a single scalar shear-diffusion equation. In the far field the two
reduce to Reynolds-free Grosch--Salwen companion systems. Their continuous spectra, the essential
spectrum of the layer operator with the translational neutral mode embedded at $\omega=0$, follow
both from the discriminants of the companion matrices and from exact dispersion relations for plane
waves $\mathrm e^{\mathrm i\xi x}$, namely $\omega=u_\infty\xi-\mathrm i(\mu_\infty/\rho_\infty)(\xi^2+\kt^2)$
for the V branch and the quartic $\det\mathbf A(\xi,\omega;\kt)=0$ for the AE branch.

The theory was verified numerically in four ways. An unrotated $9\times9$ companion
system reproduces the union of the rotated AE and V spectra for any split of $\kt$ between $k_y$ and
$k_z$, which tests the covariance theorem itself. The discriminant maps are independent of $\Rey$ to machine
precision when the outer problem is rescaled from $\Rey=1,100,10^4$. The exact dispersion curves, compared in appendix~\ref{app:dispersion-check}, trace the valleys of the
discriminant maps and give the continuous spectrum exactly. Full-spectrum Chebyshev collocation of the full, $x$-dependent V and AE operators, described in
appendix~\ref{app:evp-validation}, recovers both far-field spectra only approximately, as a dense boundary-localized cluster of least-damped eigenvalues.

The rotational reduction is the exact, Reynolds-free counterpart of Squire's transformation. Increasing the transverse wavenumber shifts the V continuous spectrum away from the
real axis exactly as its square, with every discrete vortical eigenvalue bounded below its downstream
vertex. The AE continuous spectrum is proved strictly damped by a Chu-type disturbance-energy identity.
Its least-damped point is non-increasing in $\kt$ on the evidence of a dense numerical sweep, and it
saturates beyond a Mach-dependent crossover wavenumber at the finite plateau
$-3\rho_\infty T_\infty/(4\gamma M_1^2\mu_\infty)$, independent of $\kt$.
Three-dimensional disturbances are therefore never less damped than the two-dimensional one with the
same in-plane wavenumber, in terms of the least-damped point of the spectrum, exactly for the V branch and numerically for the AE branch. Results for
planar oblique shocks follow from the normal-shock results by a constant Doppler shift. No unstable
AE eigenvalue was found in the full-layer computations, which is numerical evidence only, since the
result of \citet{Humpherys2009} covers the one-dimensional profile.

Non-modal growth was computed in the Euclidean norm on the primitive variables, from the singular-value
decomposition of the propagator and confirmed by direct--adjoint looping. It is a property of that norm,
since the same generators are dissipative in the Chu norm and $G\equiv1$. The far-field generators are
modally stable, yet a perturbation of only $6$--$11\%$ of the generator, amplitude decreasing with increasing $M_1$ from $1.25$
to $1.75$, moves an eigenvalue into the unstable half-plane. At $k_y=1$ the optimal growth $G_{\max}$ is weak in the
upstream gas, $1.06$ to $1.20$, and substantially stronger in the downstream gas, $1.27$ to $3.43$, for $M_1$
from $1.25$ to $2$, and, as shown in appendix~\ref{app:origin-boundedness}, it remains bounded at the degenerate origin of the wavenumber plane. In the full shock layer, with a window tied to the shock thickness,
$G_{\max}$ lies between the two uniform media and strictly below the downstream one at every Mach number, so
the viscous shock structure supplies no transient amplification beyond that of the post-shock gas. Its optimal
disturbance ripples the density on the downstream flank without displacing the shock front.

Overall, at low Mach number, weak shocks are strongly dissipative, with a damped
spectrum and only modest non-modal growth. The far-field spectra of \S\ref{sec:transient-results} add a further observation.
At $k_y=1$ the acoustic eigenvalues are purely imaginary, i.e.\ overdamped, in the upstream gas but form an oscillatory, weakly damped complex-conjugate pair
in the downstream gas, because $\nu k_y/c$ exceeds a critical value close to unity upstream and falls below it downstream. The shock layer thus separates a medium that cannot sustain transverse sound waves
from one that can, an asymmetry that may bear on the receptivity of the layer to disturbances, in particular those originating downstream. Since
$\nu k_y/c\approx1$ corresponds to wavelengths of the order of the mean free path, whether this asymmetry persists beyond the
Navier--Stokes--Fourier closure is a question for the kinetic effects mentioned below. Many questions remain open before this viscous picture can be reconciled with the modal
resonances found by inviscid analyses such as that of \citet{MilmanKarp2026}, among them the coupling of the shock layer to a
downstream body and the kinetic effects that lie beyond the Navier--Stokes--Fourier closure as the Mach number increases.

\begin{bmhead}[Acknowledgements.]
Work supported by Office of Naval Research Grant No.\ N00014-23-1-2839
"Kinetic Treatment of Sources and Mechanisms that Drive Unsteady Shock-dominated Flow Instability"
with Dr.\ Eric Marineau as Program Officer.
\end{bmhead}

\begin{bmhead}[Declaration of Interests.]
The author reports no conflict of interest.
\end{bmhead}

\begin{bmhead}[Data availability statement.]
The Fortran codes used for the base flow, the full-layer
eigenvalue problems and the transient-growth computations, and the scripts
that build the far-field companion matrices $\mathbf M_{\rm AE}$, $\mathbf M_V$, the
discriminant maps and the grid-refinement check of \S\ref{sec:results}, as well as the scripts that 
generated the figures presented, alongside the corresponding data files, are available 
from the author on request.
\end{bmhead}

\bibliographystyle{jfmdoi}
\bibliography{references}

\clearpage
\appendix
\section{Full-layer verification of the vortical and acoustic--entropy operators}
\label{app:evp-validation}

\begin{figure}[!b]
\centering
\includegraphics[width=0.46\textwidth]{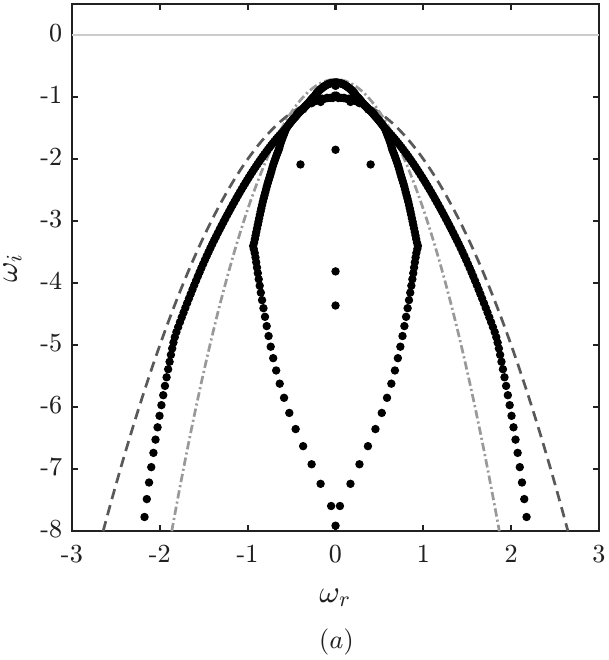}\hfill
\includegraphics[width=0.46\textwidth]{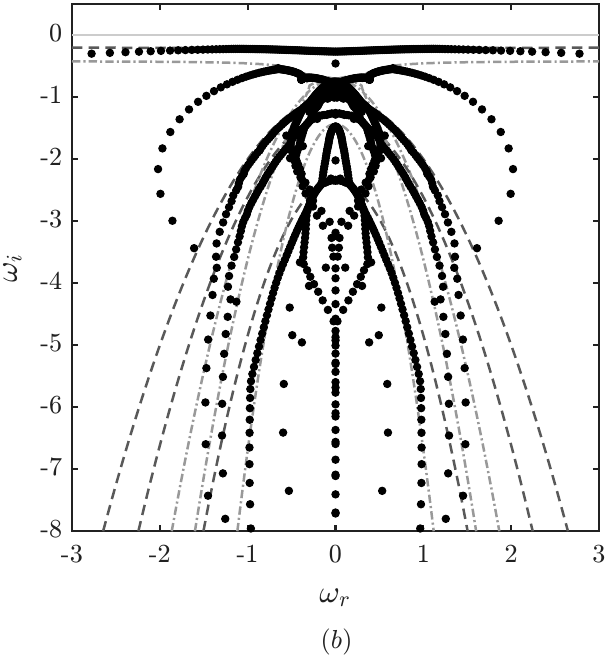}
\caption{(a) V branch, (b) AE branch at $(M_1,\gamma,\Pran,k_t)=(1.5,5/3,2/3,1)$. Dots,
full shock layer ($N=600$ and $350$, $L_x=300$); dashed, cold-side and dash-dotted, hot-side
analytic dispersion curves.}
\label{fig:evp-full-check}
\end{figure}

The far-field analysis of the main text treats each side of the shock as a uniform state.
To check it against the full layer, and to set up the discretization used for the
transient-growth analysis of \S\ref{sec:pseudospectra}, the vortical equation \eqref{eq:vorticity-Xspace} and the acoustic--entropy
system \eqref{eq:lnse-cont}--\eqref{eq:lnse-energy}, specialized as the covariance theorem allows, are discretized over the real, $x$-dependent
base flow. The method is full-spectrum computation using Chebyshev collocation on $x\in[-L_x,L_x]$ with $N$ Chebyshev--Gauss--Lobatto points,
and homogeneous Dirichlet conditions on $\wh$ (V branch) and on $\uh,\vpar,\Th$ (AE branch) at both ends.
The first-order continuity equation can carry only one condition, the inflow one, $\rh(-L_x)=0$, as in \citet{Theofilis2026},
because a condition at both ends would over-constrain this hyperbolic field and produce spurious eigenvalues with $\omega_i>0$.
Figure~\ref{fig:evp-full-check} compares the discrete spectra, at
$(M_1,\gamma,\Pran,k_t)=(1.5,5/3,2/3,1)$ with $N=600$ (V) or $350$ (AE) and $L_x=300$,
with \eqref{eq:vorticity-dispersion} and \eqref{eq:ae-dispersion-det} at both
far-field states. The agreement improves monotonically with $L_x$, since the distance
from the least-damped computed eigenvalues to the analytic curve falls by a factor of
four to five between $L_x=20$--$30$ and $L_x=300$.

Two points bear on the transient-growth results. First, the $\sim\!190$ least-damped V eigenvalues form a dense cluster
localized at $x=+L_x$, which discretizes the near-marginal vertex of the hot-side continuous spectrum.
Such a cluster is the expected signature of Dirichlet truncation of a continuous spectrum rather than a numerical defect,
and it rules out a small modal basis, which is why the propagator-norm approach of \S\ref{sec:pseudospectra} is used. Second, the least-damped AE eigenfunction is dominated by $\rh$,
in line with the density-dominated optimal disturbances of \S\ref{sec:transient-results}.

\medskip
\noindent\emph{Window and convergence of the full-layer transient growth.}
Initial disturbances and response are restricted to a window by a cosine taper, since a hard cutoff would only move the
outflow-boundary artifact to the edge of the window. The taper is $\phi=1$ for $|x|<c_1$, $\phi=0$ for $|x|>c_2$
and $\phi=\tfrac12\{1+\cos[\pi(|x|-c_1)/(c_2-c_1)]\}$ in between, and $\mathbf W={\rm diag}(\phi)$ is applied in \eqref{eq:transient-growth-def}. A wide window, $(c_1,c_2)=(60,150)$, lies well clear of the shock structure
($|x|\lesssim30$) and gives converged results, with $G_{\max}$ changing by $0.2\%$ between $L_x=300$ and $450$ and by a few percent across $N=250$--$500$.
This window reproduces the downstream far-field optimum to within $0.1\%$ at the six Mach numbers of \S\ref{sec:transient-results}, because it contains some ninety
viscous lengths of post-shock gas against a shock layer only about fifteen thick. The shock-tied window of \S\ref{sec:transient-results} changes
$G_{\max}$ by less than $0.1\%$ under threefold grid refinement near the shock. The ordering of the shock-layer, upstream and downstream values
is the same for every window and Mach number tested.

\section{Boundedness of the transient growth at the degenerate far-field origin}
\label{app:origin-boundedness}

Every entry of the matrix \eqref{eq:ae-dispersion-matrix-explicit} carries a factor of $\xi$ or $k_y$, so
\begin{equation}
\mathbf A_0(0,0)=\mathbf 0\quad\Longrightarrow\quad\mathbf A_{\rm gen}(0,0)=\mathbf 0.
\label{eq:ae-origin-zero}
\end{equation}
This vanishing reflects spatial homogeneity rather than stability, yet an unconstrained search for the optimum of
\eqref{eq:transient-growth-def} over the $(\xi,k_y)$-plane drifts toward this point, with the optimal horizon growing without
bound, which suggests a divergence. However, no divergence occurs, because along a ray $(\xi,k_y)=r(\cos\theta,\sin\theta)$ one has $\mathbf A_0=r\,\mathbf L(\theta)+r^2\mathbf Q(\theta)$
exactly, where $\mathbf L$ is the inviscid part (the symbol of the Euler equations) and $\mathbf Q$ collects the viscous terms.
With $\tau\equiv rt$ held fixed as $r\to0$,
$G\to g(\theta)\equiv\sup_{\tau\ge0}\|\exp(\tau\mathbf G_1(\theta))\|_2^2$ with $\mathbf G_1\equiv-\mathrm i\mathbf C^{-1}\mathbf L$, so the
near-origin growth is governed by the inviscid operator alone. Because the Euler system is hyperbolic, $\mathbf G_1$ has the purely imaginary
spectrum $\{0,0,\pm\mathrm ic(\theta)\}$ with $c$ real, and the double zero eigenvalue is semisimple
($\operatorname{rank}\mathbf G_1=2$ at every $\theta$ tested). Hence $\exp(\tau\mathbf G_1)$ is exactly periodic in $\tau$ and $g(\theta)$ is finite.
Over a $0.5^\circ$ scan in $\theta$, the worst case is $\sup_\theta g(\theta)=2.33$, $3.31$, $4.48$ in the upstream (cold) state and $3.36$, $6.06$, $9.60$ in the downstream (hot) state, at $M_1=1.25$, $1.5$, $1.75$.
A search of $G(\xi,k_y,t)$ over the whole plane confirms that these are the global suprema, approached but never exceeded as
$(\xi,k_y)\to(0,0)$, so that $G$ is bounded on the plane.

\section{Verification of the far-field dispersion relations against the discriminant maps}
\label{app:dispersion-check}

\begin{figure}[!h]
\centering
\includegraphics[width=0.46\textwidth]{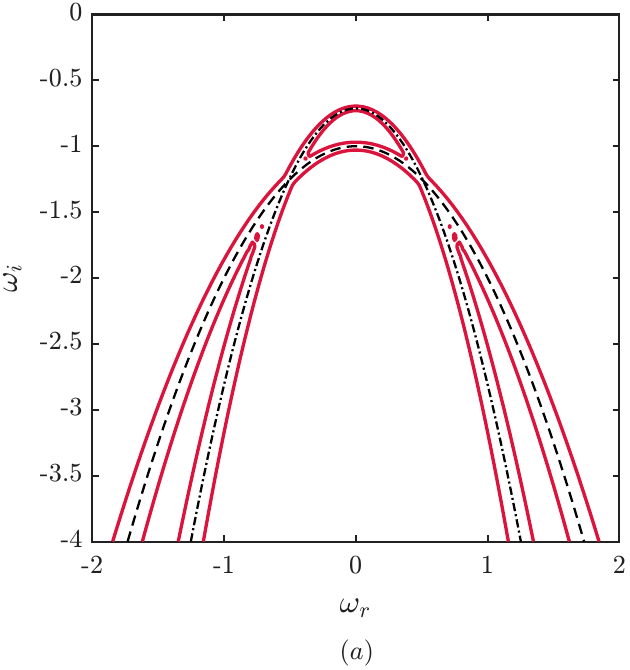}\hfill
\includegraphics[width=0.46\textwidth]{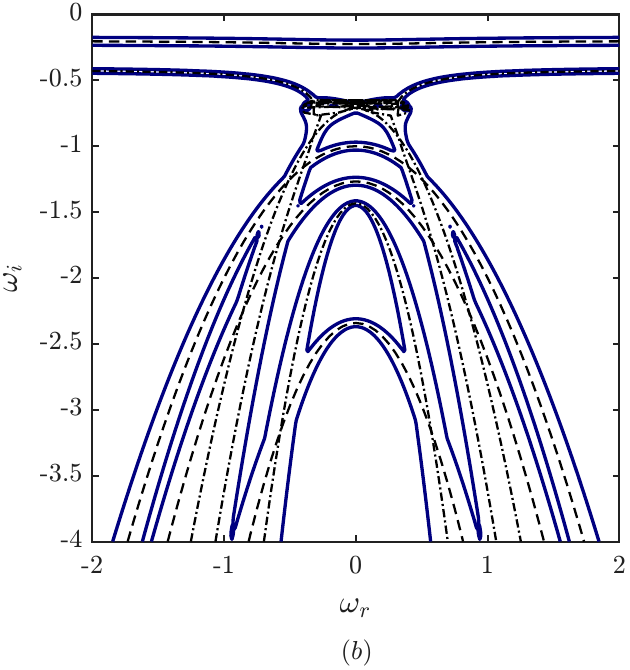}
\caption{(a) V branch, (b) AE branch, at $k_y=1$. Solid, $d_V=0.03$ (red) and $d_{\rm AE}=0.03$ (blue) contours of figure~\ref{fig:squire}(a); dashed, cold-side and dash-dotted, hot-side analytic curves.}
\label{fig:dispersion-check}
\end{figure}

As a parameter-free check of \eqref{eq:vorticity-dispersion} against the
numerical machinery of \S\ref{sec:theory-companion-discriminant}, panel (a) of figure~\ref{fig:dispersion-check}
overlays the two analytic curves $\omega_{\rm cold}(\xi,k_y)$ and $\omega_{\rm
hot}(\xi,k_y)$, $\xi\in\mathbb{R}$, on the $d_V=0.03$ contour of figure~\ref{fig:squire}(a), which has $k_y=1$.
That contour merges both far-field states \eqref{eq:discriminant}, and the analytic curves run along its two component branches.
Panel (b) does the same for the acoustic--entropy branch. It overlays the roots $\omega_j(\xi,k_y)$, $j=1,\dots,4$, of the generalized
eigenvalue problem at both far-field states on the $d_{\rm AE}=0.03$ contour of figure~\ref{fig:squire}(a). This comparison does not rely on
the companion-matrix diagonalization, so it tests the acoustic--entropy matrix implementation itself.
The collapse check of \S\ref{sec:theory-reynolds-collapse} cannot do this, since it compares two derivations built on the same implementation.
The two routes also agree to machine precision, since the discriminant evaluated at points of the analytic curves vanishes to round-off,
$d_{\rm AE}<10^{-13}$ for $\xi\in[-6,6]$, at $M_1=1.1$, $1.5$ and $2$, for both far-field states and $\kt=1,\sqrt2,2$.
The vortical discriminant is likewise zero identically, $d_V=0$, because $\sigma=\mathrm i\xi$ solves \eqref{eq:vorticity-sigma-quadratic}.

\end{document}